\documentclass[structabstract]{aa}
\usepackage{txfonts}
\usepackage{graphicx}
\usepackage{epsfig}
\usepackage{natbib}
\usepackage{amstext,cases}
\bibpunct{(}{)}{;}{a}{}{,}

\begin{document}

\title{On emission-line spectra obtained from evolutionary synthesis models} 
\subtitle{II. Scale-relations and the estimation of mass dependences}

\author{Marcos Villaverde\inst{1} \and Miguel Cervi\~no\inst{1} \and
  Valentina Luridiana\inst{2,3,1}}

\institute{Instituto de Astrof{\'\i}sica de Andaluc{\'\i}a (CSIC),
  Glorieta de la Astronom{\'\i}a s/n, 18080 Granada, Spain \and
  Instituto de Astrof{\'\i}sica de Canarias, C/ V{\'\i}a L{a}ctea s/n,
  38205 La Laguna, Spain \and Departamento de Astrof{\'\i}sica,
  Universidad de La Laguna (ULL), E-38205 La Laguna, Tenerife, Spain }

\offprints{M. Villaverde, M. Cervi\~no, V. Luridiana;
\email{mva@iaa.es, mcs@iaa.es, vale@iaa.es}}
\date{Received 20 May 2009; accepted 6 May 2010}

\abstract 
{}
{In this paper we study the influence of the ionizing cluster mass on the emission line spectrum of \ion H{ii} regions in order to determine the influence of low mass clusters on the integrated emission line spectra of galaxies.} 
{For this purpose, we present a grid of photoionization models that covers metallicities from $Z=0.001$ to $Z=0.040$, ages from 0.1 to 10 Ma (with a time step of 0.1 Ma), and cluster initial masses from 1 to 10$^7$ M$_\odot$. The stellar masses follow a Salpeter initial mass function (IMF) in an instantaneous burst mode of star formation. We obtain power-law scale-relations between emission-line luminosities and ionizing cluster masses from the grids and we evaluate the dependences on the ionizing cluster mass for some line luminosities, equivalent widths and line ratios.} 
{Power-law scale-relations are shown to be useful tools to obtain robust diagnostics, as examples: (a) H$\alpha$/H$\beta$ ratio varies from the usually assumed value of 2.86, these variations imply the existence of a lower limit to the attainable precision in extinction estimations of $\Delta E(B-V) \sim 0.1$.; (b) $EW$(H$\beta$) is a good age indicator with a small dependence on cluster mass, while $EW$([\ion O{iii}] 5007) shows a noteworthy mass dependence; (c) abundance estimations from $R_{23}$ are practically unaffected by variations of the cluster mass; (d) estimations from $S_{23}$ and $\eta^\prime$ would improve if the cluster mass dependences were considered and (e) [\ion O{ii}] 3727/H$\alpha$ is a good star formation rate indicator for ages older than $\sim$4.5 Ma. We also show that the ionizing cluster mass dependence explains why empirical calibrations produce more reliable diagnostics of some emission lines than photoionization models grids. Finally, we show preliminary results about the contribution of low mass clusters ($M < 10^4$ M$_\odot$) to the integrated emission line spectra of galaxies, which can be as high as 80\% for some relevant lines.} 
{}
\keywords{HII regions -- Galaxies: star clusters -- Galaxies: stellar populations}

\titlerunning{Emission-line spectra and synthesis models (II)}

\maketitle

\section{Introduction and motivation}\label{sec:introduction}

The intrinsic properties of star forming regions are often inferred
from observations of emission-line indicators. Examples of these
indicators are the equivalent width of H$\beta$ ($EW$(H$\beta$)) for
age estimations, the radiation softness parameter $\eta\prime$
\citep{VP88} for cluster effective temperature ($T_{\mathrm{eff}}$)
estimations, or the $R_{23}$ (also called O$_{23}$) \citep{Petal79}
and $S_{23}$ \citep{VE96} parameters as abundances indicators. Some
reviews that deal with these indicators have been written by
\citet{F03}, \citet{Sta04} and \citet{Sta07}.

These and other indicators are used in estimation-methods and/or
classification methods, like the \citet{BPT} and \citet{VO87}
diagnostic diagrams. These methods make use of only a handful of
emission lines, more specifically, H$\alpha$, H$\beta$ and some
collisional lines of ions of O, N and S, so the understanding of the
dependence of the intensities of such lines on the physical properties
would allow us to calibrate the estimation methods.

When using theoretical methods for such indicator-calibration
purposes, photoionization models and stellar population synthesis
models prove to be very useful tools, as they allow the control of
relevant physical parameters and return the possible emission line
spectrum associated with such physical conditions. Nevertheless, one
of the main issues that limits their power is the difficulty of
establishing whether the solution found is unique for a given
observation. One way to deal with such problem is building tailored
models in which as many observational constraints as possible are
assumed: an example of this strategy is offered by \cite{LPL99,LP01}
and \cite{LPPC03}, where models that simultaneously reproduce the
intensity of several emission lines observed through different
apertures are computed. A different approach is to compute grids or
sequences of models. In this case fewer parameters are used but they
span a larger range of values; this allow to apply the grids or
sequences to diverse objects. Some examples can be found in the works
of \citet{Huch77}, \citet{Olofsson}, \citet{GVD94}, \citet{SI03} and
\citet{Detal06}, among others.

One of the parameters that can be considered in the model grids for
star forming regions is the ionizing cluster mass. Yet, in spite of
being a fundamental physical property, little work has been done
regarding the ionizing cluster mass. Examples can be found in
\citet{GVBD95a,GVBD95b}, \citet{SL96}, \citet{SSL01}, and in a more
indirect way in \citet{Detal06}, nevertheless the explicit dependence
with the mass is not discussed in these works.

In this work we center our attention explicitly on the influence of
the ionizing-cluster mass on the emission-line spectra. There are two
motivations for this choice. The first one is the long term goal of
investigating the accuracy of the modeling of the emission line
spectra of ionizing clusters by means of evolutionary synthesis
models, and the influence of sampling effects in the initial mass
function (IMF) on the emission line spectrum \citep[ Paper
I]{Cetal03}. In the first paper of the series we investigated when
sampling effects in the stellar populations of the cluster cannot be
neglected in the modelling of the emission line spectrum, and we
showed that cluster masses must be larger than $10^3$ M$_\odot$ to
obtain realistic results for relevant emission lines, $10^3$ M$_\odot$
being the cluster mass limit where a single star is as luminous as the
whole cluster \cite[that is, the {\it Lowest Luminosity Limit} (LLL),
see also][~for more details]{CL04}. \cite{CL04} propose to use the
mean values obtained from synthesis models in clusters at least 10
times larger than the LLL to minimize IMF sampling effects. In this
second paper we aim to obtain a reference scale (calibrated with the
cluster mass) that provide a mean reference value that allows to
evaluate quantitatively the dispersion due to the IMF sampling. As it
is known, and we will see below, the intensity of hydrogen
recombination lines roughly scale with the number of ionizing photons,
and hence with the number of massive stars and the mass of the
cluster. However, a similar scale relation has not been established
for the case of forbidden lines. In the last paper of this series, we
will use these scale relations to evaluate the variance due to IMF
sampling in the emission line spectrum as a function of the cluster
mass.

The second motivation is to investigate the emission line spectrum of
galaxies by the composition of different H{\sc ii} regions. In
general, the emission lines spectrum of galaxies has been modeled by
means of single photoionization models where all the stars are
concentrated in a single point (whatever the stellar density of such
hypothetic point) and with a single ionized gas nebula \cite[ as
examples]{SSL01,ChL01}. Only few attempts have been done to consider
an ensemble of clusters \cite[][]{JM08} and to include an initial
cluster mass function (ICMF) in the simulations \cite[][]{Detal06b}.
It can be claimed that the form of the ICMF\footnote{Assuming both a
  power law with a slope of $-2$ \cite[][]{LL03}, or a log normal with
  $\sigma=1.7$ and $M_0=100$ \cite[][]{Detal06}.} is such that each
logarithmic bin of cluster mass contributes a similar amount to the
total flux and massive clusters will therefore tend to wash out the
stochastic effects of low-mass clusters \cite[][]{Detal06}. However,
this statement implicitly assumes that the flux of {\it all relevant
 emission lines, including collisional ones}, scale linearly with the
cluster mass. Checking the validity of such statement is the second
objective of this paper.

To study these questions we have computed a grid of photoionization
models with ionizing clusters of different masses. In this paper we
do not consider sampling effects but the scale-relations of the
emission line intensities with the ionizing cluster total mass. We
have included in our study some extreme (unrealistic) cases of cluster
masses (e.g. clusters with masses lower than $10^3$ M$_\odot$,
affected by sampling effects, or clusters with masses larger than
$10^7$ M$_\odot$, which are not consistent with the assumption of an
instantaneous burst scenario). Such extremes allow us to explore the
range where the scale relations are valid in a photoionization model,
despite that they cannot represent real clusters, as well as to check
if the emission line spectrum of a non-active galaxy can be modeled as
a single nebula.

In the next section the properties of this grid of photoionization
models are described and the differences with grids from other authors
are discussed. In section \ref{sec:results} the results of the models
are analysed and the luminosity functions are computed. In
subsections \ref{subsubsec:diagdiag} and \ref{subsubsec:parameters}
the reliability of these functions is checked comparing it with the
results obtained directly from the models. The implications that arise
from the analysis are discussed in section \ref{sec:discussion}, and
finally the conclusions of this work are summarized in section
\ref{sec:conclusion}.
 
\section{Modeling strategy: defining the grid}
\label{sec:model}

In the following we describe the input parameters assumed for the
grid, and the differences with the grids elaborated by other authors.

\subsection{Continuum shape and intensity}

In this work we have used the spectral energy distributions (SEDs)
computed with the evolutionary code by \cite{CMHK02}. All the
computations assume an instantaneous burst of star formation and a
\cite{sal55} IMF slope in the mass range 2 to 120 M$_\odot$. We have
adopted the evolutionary tracks with standard mass-loss rates by
\cite{Schetal92,Schetal93,Schetal93b,Charetal93}, and the following
atmosphere models: \cite{CoStar} {\sc CoStar} for main-sequence hot
stars more massive than 20 M$_\odot$, \cite{Schmetal92} for WR stars,
and \cite{kur} for the remaining stars. No X-ray emission due to the
conversion of kinetic energy or the presence of young Supernova
Remnants has been included. The models have been computed for the
five metallicities available in the Geneva track dataset ($Z$=0.001,
0.004, 0.008, 0.020 and 0.040) and cover the age range from 0.1 to 10
Ma with a time step of 0.1 Ma. The resulting SEDs have been used as
an input to the photoionization code {\it Cloudy} version 08.00, last
described by \citet{Fetal98}.

The continuum intensity has been defined in terms of the amount of
photons able to ionize hydrogen, $Q(\mathrm{H}^0)$. We have used
$Q(\mathrm{H}^0)$ values that correspond to amounts of gas transformed
into stars at the onset of the burst ranging from 1 to 10$^7$
M$_\odot$. The mass step in the most relevant range ( $10^3$- $10^6$
M$_\odot$) is 0.5 dex while for lower masses the mass step is 1 dex.
Obviously the 1 M$_\odot$ case is just an academic case, since the
lower mass limit of the IMF is 2 M$_\odot$. Moreover, clusters with
masses below $10^3$ M$_\odot$ are also academic cases, since the
(average) SED used would not be representative of real clusters
\citep{Cetal03, CVG03, CL04, CL06}. Whatever the case, the extension
down to fictitious 1 M$_\odot$ clusters allows to explore the
proportionality of the resulting emission line spectrum with the
incident ionizing flux, independently of how "real" the incident flux
is.

This strategy differs from the one by \cite{Dop00}, who use the SED of
a zero-age stellar cluster to build a grid where the metallicity and
the ionization parameter of the model nebulae are varied. It is
important to note that using a zero-age SED is roughly equivalent to
using a single stellar atmosphere model with $T_{\mathrm {eff}}$
between 45,000 and 50,000 K \cite[c.f.][]{MHK91,GVBD95a}, with the
exact value depending on metallicity. One of the conclusions of these
authors, specifically that H{\sc ii} regions can be modelled with
zero-age SEDs, contrasts with the philosophy underlying evolutionary
synthesis models, which are in their essence {\it evolutionary}.
Additionally, were the SED of observed H{\sc ii} regions described by
synthesis models at zero-age, then all observed H{\sc ii} regions
should have similar H$\beta$ equivalent widths (with values around 500
\AA), which is clearly in contradiction with the observations. The
fact that their grid reproduce the observed trends in some diagnostic
diagrams is a nice example of the non-uniqueness of the possible
physical conditions that produce a given emission line spectrum.

A strategy more similar to ours is the one by \cite{GVBD95a,GVBD95b},
who present their results in terms of ages, metallicity and initial
amount of gas transformed into stars. Our modelling differs from
theirs in (a) the metal mixture, especially for what concerns the N/O
ratio, which in our grid varies with the metallicity, and (b) the
geometry, which is spherical in our case and plane-parallel geometry
in \cite{GVBD95a}.

Our grid is similar in approach to the one by \citet{SL96} and
\citet{SSL01}. Differences are in this case the metal mixture, and the
higher resolution in age and cluster size in our grid with respect to
theirs.

Finally, \citet{Detal06} used for the grid a parameterization on
metallicity, age and on a parameter that depends both on the ionizing
cluster mass and on the pressure in the interstellar medium (ISM). In
this sense the cluster mass is considered but its effects are mixed
with the ones of the pressure. There are also differences in the metal
mixture, in the age covering, which in our work is larger and has
higher resolution, and in the covering factor , which in their work
varies with the age.

\subsection{The chemical composition mesh}

In this work we assumed a dust-free nebula and the solar composition
listed in Hazy \citep[][ table 14]{hazy}, which corresponds to the
abundances by \cite{GA89} with the extensions by \cite{GN93} except
for C, N, O and He. The solar oxygen abundance has been set to $O /H$
= $4.91 \times 10^{-4}$, following \cite{APLA01}. The solar C/O ratio
has been set to 0.5 \citep{APLA02}, yielding a solar C abundance of
$2.45 \times 10^{-4}$. For metallicities other than solar, the
abundances have been linearly scaled with $Z$ except for He and N.
For the He abundance, we have used the following relation:

\begin{eqnarray}
\frac{He}{H} & = & 0.0786 + 18.4 \left(\frac{ O}{ H} \right),
\end{eqnarray}

\noindent which has been obtained from the value of the primordial
helium abundance obtained by \cite{LPPC03}, $Y_{\mathrm p}$ = 0.2391,
and the slope of the He vs. O relationship quoted by \cite{PPR00},
${\Delta Y}/{\Delta \mathrm O}$ = 3.5. \citet{Detal06} assume for the
He abundance a steeper relation given by:

\begin{eqnarray}
\frac{{He}}{ H} & = & 0.0737 +52.5 \left(\frac{ O}{ H} \right).
\end{eqnarray}

In the case of nitrogen, we have assumed the $N/O$ ratio given by:

\begin{tabular}{lclr}
&&&\\
$\log \left(\frac{ N}{ O}\right)$ & = & $-1.5 $ & 
 for $\log \left(\frac{ O}{ H}\right) < -4. $\\
  & = & $1.1 \log \left(\frac{ O}{ H}\right) +2.9 $ & 
 for $\log \left(\frac{O}{ H}\right) > -4.$ \\ 
&&&\\
\end{tabular}

The relation has been obtained by a fit by eye of the data presented
in \cite{PTV03}. It should be noted that, as shown by \cite{PTV03}, an
exact $N/O$ ratio does not exist, since its value changes from system
to system even for a fixed metallicity. This dispersion could arise
naturally if the chemical evolution of a galaxy were driven by small
star forming regions, so that the sampling effects of the IMF in the
chemical evolution become relevant \citep{WA83,WK95,CM02,CH08}. Given
this situation, we think it is more meaningful to fit the results by
eye rather than using a formal method.

We have also performed a consistency check on the assumed abundances
of the stellar tracks. As explained in the corresponding papers
\citep{Schetal92,Schetal93,Schetal93b,Charetal93}, the tracks assume a
helium content by mass given by:

\begin{eqnarray}
Y= & = & Y_{\mathrm p} + \frac{\Delta Y}{\Delta
Z} Z,
\end{eqnarray}

\noindent with $Y_{\mathrm p}=0.24$ and ${\Delta Y}/{\Delta Z}=3$ for
all metallicities except for the case $Z$=0.040 \cite[ twice
solar]{Schetal93} where ${\Delta Y}/{\Delta Z}$ takes a value of 2.5.
Additionally, the relative ratios of heavy elements follow the ones
obtained by the \cite{AG89} solar mixture, somewhat different from the
values by \cite{GA89} quoted before, for all the metallicities. The
absolute values have been scaled with $Z$ except for $Z$=0.004
\citep{Charetal93} and $Z$=0.040 \citep{Schetal93} where
\cite{G91}\footnote{The reference paper cites \cite{Getal91} instead
 of this one.} has been used. Under this assumption, a $N/O$ ratio
of 0.15 is assumed for all the metallicities except for $Z$=0.004 and
$Z$=0.040, where it is 0.13. This is clearly inconsistent with the
mixture we assume in the nebular gas, but, since we are not able to
quantify the influence of a change of the $N/O$ ratio on stellar
evolution, and we prefer to adopt a nebular metal mixture more
consistent with the observational data, there is no solution other
than maintaining the inconsistency and hoping that it only produces
minor effects.

Finally, we have obtained the oxygen abundance from the different set
of tracks. The resulting $O / H$ values for each metallicity, $Z$, are
summarized in table \ref{tab1}, where we also show the gas $O / H$ and
$Z$ values assumed in \cite{SSL01} for comparison.

\begin{table}
\begin{tabular}{c|ccc}
\hline
Z&$O /H$ & Stasi{\' n}ska, Schaerer, & This work\\
&(tracks)& \& Leitherer (2001) & \\
\hline
0.0004 &  --                & $1.70 \times 10^{-5}$&  --                 \\
0.001  & $4.39\times 10^{-5}$ & --                   & $2.46 \times 10^{-5}$\\
0.004  & $1.77\times 10^{-4}$ & $1.70 \times 10^{-4}$& $9.82 \times 10^{-5}$\\
0.008  & $3.64\times 10^{-4}$ & $3.40 \times 10^{-4}$& $1.96 \times 10^{-4}$\\
0.020  & $9.75\times 10^{-4}$ & $8.51 \times 10^{-4}$& $4.91 \times 10^{-4}$\\
0.040  & $2.12\times 10^{-3}$ & $1.70 \times 10^{-3}$& $9.82 \times
10^{-4}$\\
\hline
\end{tabular}
\caption[]{Values of $O /H$ implicit in the evolutionary tracks, in \cite{SSL01}
grid and in this work for a given $Z$.}
\label{tab1}
\end{table}

Since there is no feasible way of solving the inconsistency between
the abundances of the evolutionary tracks and those adopted for the
gas, present in our and similar works, we limit to call the attention
to it and present our model grid in terms of $Z$, without expressing
the metallicity as a function of the solar value. It should be kept
in mind that a direct comparison with other works must be done with
extreme caution.

\subsection{Density, geometry, filling factor and covering factors}

The assumptions on the electronic density, the geometry, the filling
factor and the covering factor define the properties of the model
nebulae. For the case of the density and geometry we have assumed
rather conservative values: a hydrogen constant density of 100
cm$^{-3}$ distributed homogeneously over an expanding spherical
geometry in an ionization bounded configuration. The assumed density
of 100 cm$^{-3}$ is typical for \ion H{ii} regions and it is lower
than the critical densities of the species used in this work. The
internal radius of the cloud, $R_{\mathrm in}$, has been fixed for all
the models to 2 pc ($\log R\mathrm{(cm)}=18.790426$).

The covering factor $\Omega/4\pi$ is the fraction of the solid angle
around the ionization source covered by the gas nebula. There is a
large amount of observational evidence \cite[see e.g. ][ among
others]{Betal00,ZBRR02,CDTT02,IPMT02} that a substantial fraction of
the ionizing photons escape from the ionized nebulae. This finding
may indicate either that the covering factor is smaller than one, or
that the nebula is density bounded (or a mixture of both situations).
Although the predictions of photoionization models differ in the two
situations \cite[see, e.g.][ for a full discussion]{LPPC03}, we have
chosen to model the escape of ionizing photons adopting a covering
factor and assuming the nebula ionization bounded, rather than
assuming it density bounded. This choice has been motivated by
simplicity considerations, since it offers the advantage of preserving
the proportionality between the number of ionizing photons and the
intensity of Balmer lines, in contrast to what happens in density
bounded nebulae. This means that we can test $L$(H$\beta$)
estimations by synthesis models based on $Q$(H$^0$). In this work we
have assumed a covering factor equal to 1, although estimated values
of the covering factor range between 0.2 and 0.8 \citep{ZBRR02}.
Again, our choice is based on simplicity considerations, since our
objective is to obtain the shape of the luminosity functions of the
emission lines intensities but not to calibrate it.

For each age, metallicity and mass, the photoionization model has been
computed until a minimum electron temperature has been reached. This
minimum electron temperature is 4000K for Z=0.001, 0.004 and 0.008 and
200K for Z=0.020 and Z=0.040. Once a solution has been found, it has
been used as input for another iteration, and so on, until a stable
solution has been found.

Additionally, we have obtained the continuum flux resulting from
Cloudy, obtained directly with the {\tt punch continuum} command.
However, this continuum includes the luminosity of the resulting
emission lines. To exclude the emission line contribution, we have
obtained it using the {\tt punch outward} continuum command and then
subtract it from the continuum flux \citep[see][ for more details on
the use of this command and the {\tt punch continuum} one]{hazy}. The
resulting continuum has been used later to obtain the equivalent
widths of the corresponding emission lines.

We want to note that the choices made here, especially those
concerning the geometry, density, filling factor and covering factors
are merely convenient choices: we are not suggesting that all H{\sc
 ii} regions can be modelled with this set of parameters. We are
interested here only in obtaining a rough estimation of the relation
of intensity of relevant emission lines with the stellar cluster mass
and how these dependencies vary from line to line. This goal is the
driver in our choice of the grid which drastically differs from how
other authors define their grids, since they address a different
problem from the one discussed here.

\section{Results of the grid models}
\label{sec:results}

In this section we analize the results from the models. First we
center our attention on the two hydrogen recombination lines H$\alpha$
and H$\beta$, specifically on their use for age and extinction
estimations. Next we derive mass-luminosity scale-relations from the
models and apply the same method to some relevant collisional lines.
Finally we check the scale-relations results with the ones obtained
directly from models.

\subsection{Hydrogen recombination lines}
\label{subsec:Hlines}

\subsubsection{H$\beta$ equivalent width}

\begin{figure}
  \resizebox{\hsize}{!}{\includegraphics[width=7cm]{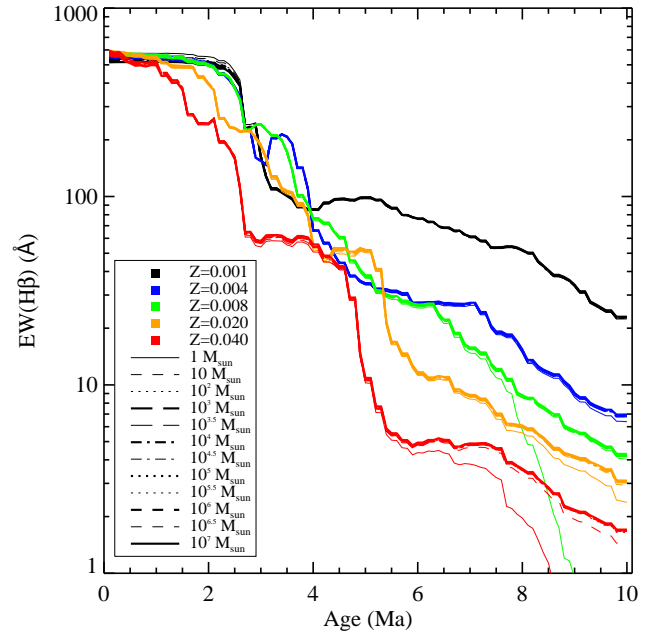}}
  \caption[]{H$\beta$ equivalent width versus age. Grey scale colors
    represent metallicities. Thin solid line: 1 M$_\odot$. Thin
    dashed line: 10 M$_\odot$. Thin dotted line: 100 M$_\odot$. Thin
    long-dashed line: 10$^3$ M$_\odot$. Thick dash-dot line: 10$^4$
    M$_\odot$. Thick dotted line: 10$^5$ M$_\odot$. Thick dashed line:
    10$^6$ M$_\odot$. Thick solid line: 10$^7$ M$_\odot$.}
  \label{fig:EWhb}
\end{figure}

The resulting $EW$(H$\beta$) from the grid is shown in Fig.
\ref{fig:EWhb}. The plot shows the results for the complete grid
(including all the assumed cluster masses). Different line types
account for cluster masses while grey scale colors represent
metallicities. For a given metallicity, the curves at different
masses are similar with just slight differences that depend on the
ionizing cluster mass. Excluding the 1 M$_\odot$ cases, these
differences are always less than 0.1 dex. The 1 M$_\odot$ cases
diverge from the rest of models at old ages because the computation
goes deeper into neutral gas and the continuum at H$\beta$ wavelenght
increases.

\begin{figure}
  \resizebox{\hsize}{!}{\includegraphics[width=7cm]{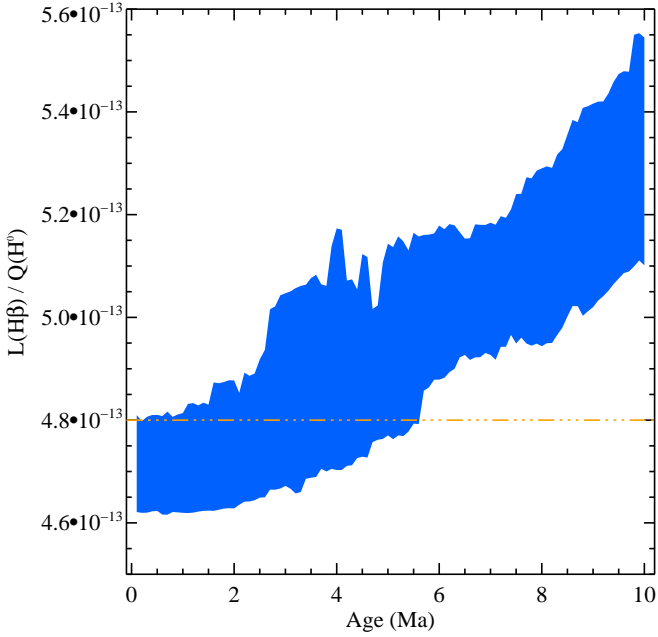}}
  \caption[]{Ratio of the H$\beta$ luminosity to the number of
    hydrogen ionizing photons, $Q$(H$^0$), versus age for models with
    $M \ge 10^3$ M$_\odot$. The shaded zone corresponds to the range
    of values of the ratio. Dash-dot-dot line: value assumed by
    synthesis models.}
\label{fig:LhbQH}
\end{figure}

As a safety check to the resulting $EW$(H$\beta$) from synthesis
models computations, we have compared the ratio of the H$\beta$
luminosity with the number of hydrogen ionizing photons, $Q$(H$^0$).
Synthesis models assume a putative constant value around
$L(\mathrm{H}\beta)= 4.8 \times 10^{-13}(\Omega/{4\pi})
Q(\mathrm{H}^0)$ \citep[e.g.][]{CMH94,LH95} for $T_{\mathrm{e}}$=
10$^4$ K, $n_\mathrm{e}$= 100 cm$^{-3}$ and He/H = 0.1. Figure
\ref{fig:LhbQH} shows that the resulting conversion value for $M >
10^3 $ M$_\odot$ only varies $\sim$20\% during the considered ages.

Summing up, $EW$(H$\beta$) is a ``good enough'' indicator of the age
of an \ion H{ii} region \citep{CPD86} once the covering factor is
known. $EW$(H$\beta$) depends mainly on the metallicity \citep{CMH94}
and age of the cluster with almost no variations due to the size of
the cluster, if sampling effects are not considered \citep[but
see][]{Cetal00,Cetal02b}.

\subsubsection{H$\alpha$, H$\beta$ and extinction}

\begin{figure}
  \resizebox{\hsize}{!}{\includegraphics[width=7cm]{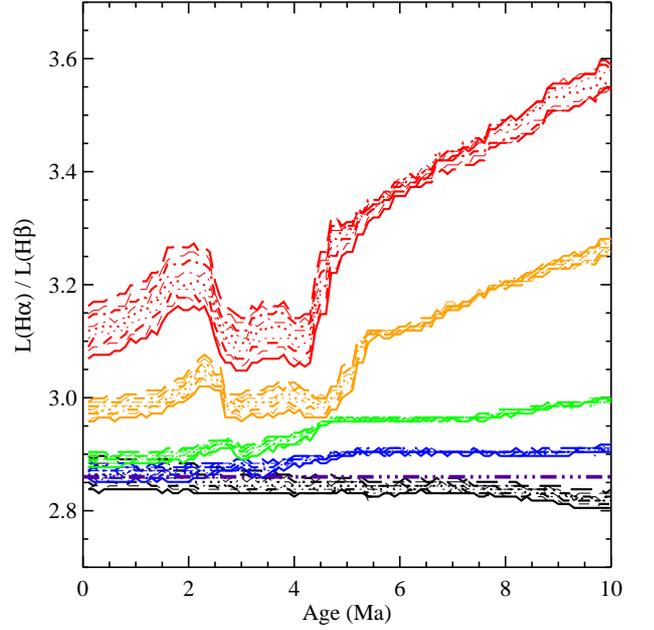}}
  \caption[]{H$\alpha$/H$\beta$ ratio versus age. Dash-dot-dot line:
    normally assumed value. Rest of symbols as in Fig.
    \ref{fig:EWhb}.}
\label{fig:Ext}
\end{figure}

More interesting is the behaviour of the H$\alpha$/H$\beta$ ratio,
used commonly as an extinction index, and for which a constant ratio
of 2.86 is normally assumed. Figure \ref{fig:Ext} shows such ratio,
obtained for models with $M \geq 10^3$M$_\odot$, the line types have
the same meaning as in Fig. \ref{fig:EWhb}. The ratio computed for
case B and $T_{\mathrm{e}}$= 10$^4$ K is also shown for comparison.
The figure demonstrates that in fact such an assumption is incorrect.
The largest discrepancies between the results of the models and the
value normally assumed occur generally for $Z$=0.040. Nevertheless,
the largest ones correspond to models with $Z$=0.008, 1 M$_\odot$ (not
shown in the figure) and ages around 10 Ma, which are $\sim$50\%
greater than the usual assumed value and lead to $E(B-V)$ = 0.36. For
that cluster mass the mean spectrum obtained by synthesis models is
just an academic result, as we pointed out before. Considering only
the models with $M \geq 10^3$M$_\odot$ a maximum discrepancy of 25.8\%
with respect to the ratio normally assumed is found for models with
$Z$=0.04, 10 Ma, and masses larger than $10^6 $M$_\odot$. However,
this only sets an upper limit to the bias since at such ages the \ion
H {ii} region are so faint that they are not detected. Considering
lower and more relevant ages (0.1-6 Ma), differences near 18\% are
found for $Z$=0.04 and masses 10$^{3.5}$ and 10$^4$ M$_\odot$ at 6 Ma.
If we also exclude the highest metallicity models, differences near
9\% are found for $Z$=0.02, $M \geq {4.5}$M$_\odot$ at ages near 6 Ma.

The use of an incorrect H$\alpha$/H$\beta$ ratio leads to an incorrect
extinction estimation which affects the line intensities inferred.
For example, a difference of 9\% with respect to the normally assumed
ratio leads to a spurious $E(B-V)$ = 0.08. Although this value is
small, this exercise shows that $E(B-V)$ estimations from the emission
line analysis (without the use of tailored models) suffer from
intrinsic uncertainties which may be as large as 0.1. This effect is
independent of the extinction due to the \ion H{ii} region dust
content. If dust effects were taken in account, then the scatter in
the H$\alpha$/H$\beta$ ratio would be different: bigger H{\sc ii}
regions would have a larger amount of dust, the dependence with the
geometry would be stronger and a tailored analysis would be needed.

\subsubsection{Scale-relations}
\label{subsubsection:scalerelations}

\begin{figure}
  \resizebox{\hsize}{!}{\includegraphics[width=7cm]{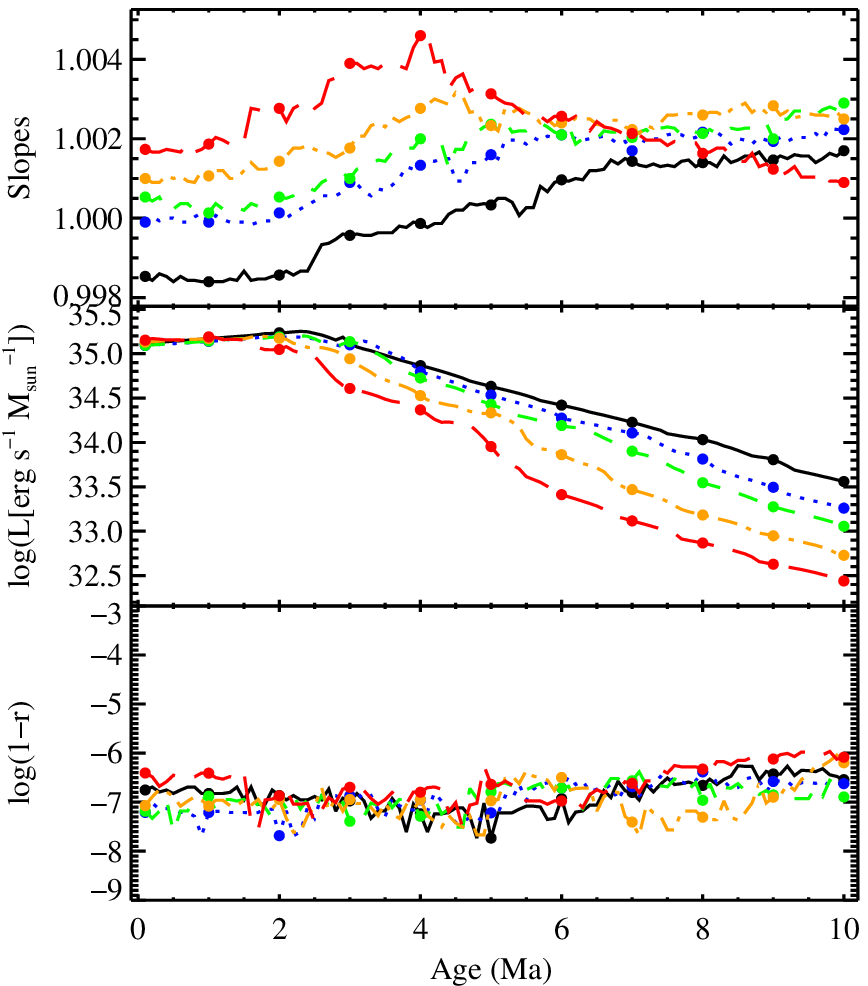}}
  \caption[]{ From top to bottom: slopes, $\log$($L$[erg s$^{-1}$
    M$_\odot^{-1}$]) and correlation coefficient of linear regression
    of $\log$($L$(H$\beta$)) versus log(M). Grey scale colors
    indicate metallicities as in Fig. \ref{fig:EWhb}. Dots indicate
    models with age equal to 0.1, 1, 2, 3, \ldots ~Ma.}
\label{fig:Hb}
\end{figure}

\begin{figure*}
  \resizebox{\hsize}{!}{\includegraphics[width=7cm]{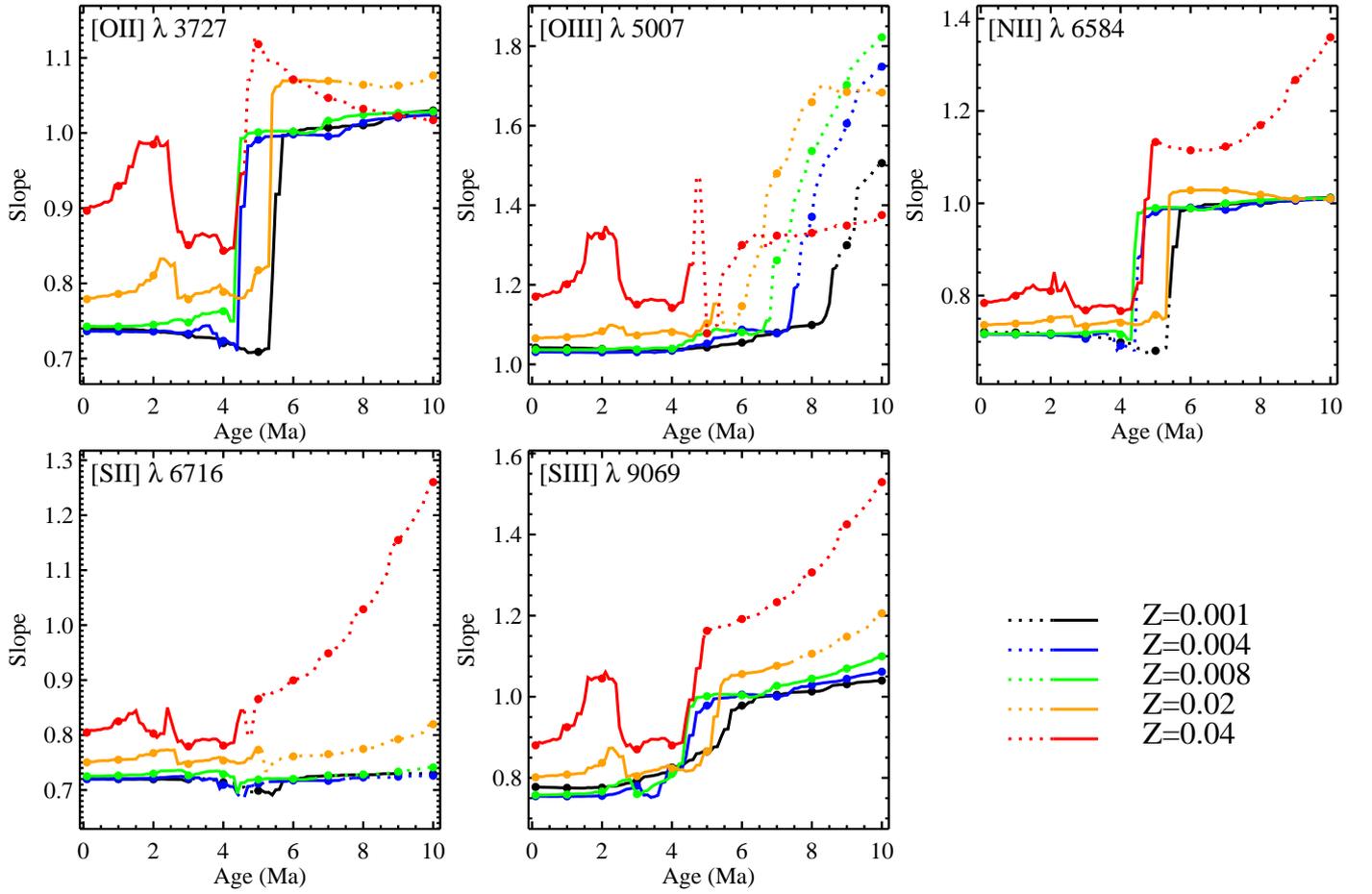}}
  \caption[]{Slopes of the scale-relations for some collisional lines.
    In the plot, the dotted section of the lines represents the ages
    where the $EW \ge 1$ \AA ~is not fulfilled for all masses in the
    $10^3 - 10^6$ M$_\odot$ mass range. Symbols as in Fig.
    \ref{fig:EWhb}.}
\label{fig:collines}
\end{figure*}

\begin{figure*}
  \resizebox{\hsize}{!}{\includegraphics[width=7cm]{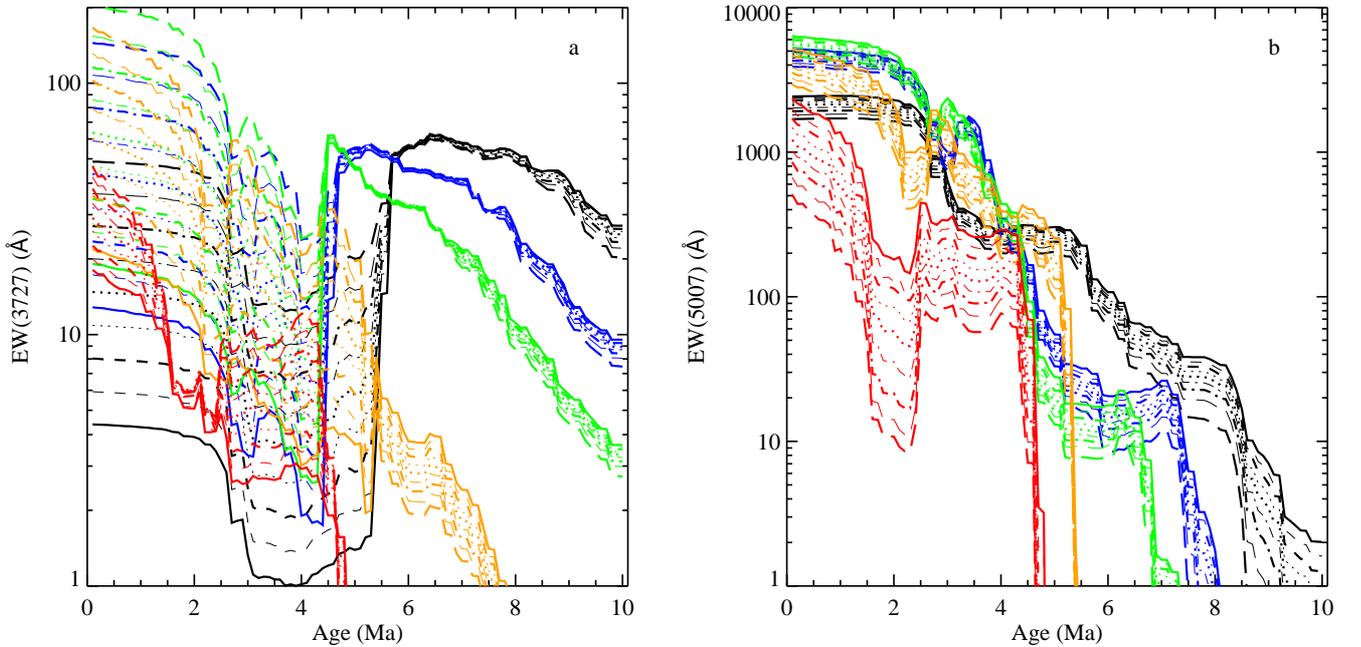}}
  \caption[]{Equivalent widths of a) [\ion O {ii}] $\lambda$ 3727 \AA
    ~\ and b) [\ion O {iii}] $\lambda$ 5007 \AA ~\ for models with $M
    \ge 10^3$ M$_\odot$. Symbols as in Fig. \ref{fig:EWhb}.}
\label{fig:EWO}
\end{figure*}

In order to study the variation of the intensities of the emission
lines considered in this work, we made ordinary least square linear
regressions of the logarithm of the line intensities versus the
logarithm of the cluster mass for each line, metallicity and age. The
linear fits provide us scale-relations between the line intensities
and the cluster mass of the form

\begin{equation}
\log (L) = \alpha + \beta\log (M), 
\label{eq:linreg}
\end{equation}

\noindent or equivalently

\begin{equation}
L  = A \times M^\beta \mathrm{ ,}
\label{eq:scalerelation}
\end{equation}

\noindent where A=10$^\alpha$. These scale-relations, as said above,
have been built to obtain a first estimation of the behaviour of the
emission lines with the mass of the cluster. Due to the assumptions
made in the models and the simplicity of the description with
power-laws, we strongly discourage from using this method when fine
results are sought. Let us stress again that our goal is just to
obtain an approximate quantitative representation of the \ion H{ii}
region luminosity function.

The regressions have been computed only with models with $M \geq
10^3$M$_\odot$. This condition to the cluster-mass agrees with the
minimum cluster mass obtained by \citet{Cetal03} for a reliable
(although affected by sampling effects) description of the ionizing
continuum by the mean ionizing continuum obtained in synthesis models.
Regressions with models in all the mass range, including the purely
academic case of clusters less massive than $10^3$M$_\odot$, have also
been computed but they have been discarded, because the agreement in
diagnostic diagrams between regressions and models in the high mass
range is worse than when only models with $M \geq 10^3$M$_\odot$ are
used (see section \ref{subsubsec:diagdiag}).

We have also studied an {\it a posteriori} additional constraint on
the observability of emission lines. This constraint has been
parameterized in terms of a distinction between lines with $EW$
smaller and larger than 1 \AA\ respectively. Of course, this is an
optimistic value: if older stellar populations were present, the
continuum level would be higher and the resulting $EW$s lower.
However, we assume just a single stellar population in an
instantaneous burst mode of star formation in our modeling. The
effects of these additional populations on the continuum level and
$EW$s measurements can be modeled through the use of a different star
formation history, which goes beyond the scope of this paper.

The $EW$ computations have been carried out using the line luminosity
and the continuum luminosity as computed by Cloudy, which takes into
account the two contributions to the continuum luminosity (stellar and
nebular continua). We have obtained continuum scale relations using
Eq. \ref{eq:linreg}. As expected, the continuum luminosity (including
the nebular contribution) scales linearly with the cluster mass,
although there is a small dispersion when the continuum obtained from
the scale relations is compared with the continuum from particular
models (around 7\%, compatible with the scatter in the hydrogen
recombination lines shown previously). We have computed the EW of the
emission line using the scale relations following:

\begin{equation}
EW = \frac{A_\mathrm{line} \times M^\beta}{A_\mathrm{cont} \times M} = \frac{A_\mathrm{line}}{A_\mathrm{cont}} \times M^{\beta - 1}, 
\label{eq:ew}
\end{equation}

\noindent where $A_\mathrm{line}$ , following Eq.
\ref{eq:scalerelation}, is the luminosity of the line when the cluster
mass is 1 M$_\odot$, obtained from the extrapolation of the power law
down to this value. Analogously, $A_\mathrm{cont}$ is the luminosity
of the continuum when the cluster mass is 1 M$_\odot$. Since we are
interested in the relation of the luminosities with the cluster mass,
we center our attention on the slopes of the scale-relations.
However, the $EW$ obtained from the scale-relations are more dependent
on the geometrical assumptions (inner radius, covering and filling
factors) than the scale-relations slopes themselves, due to the large
dependences of $A_\mathrm{line}$ and $A_\mathrm{cont}$ on the
geometry. So, we will use this constraint on $EW$ as a
self-consistency check by comparing the results from models that
fulfill $EW \ge 1$ with scale-relations results with $EW$ estimations
obtained from the scale-relations themselves. The values of the
normalized luminosities will be shown in a posterior work when a
proper ICMF and star formation history will be included for models of
non-active emission line galaxies.

In Fig. \ref{fig:Hb} the slopes, the luminosity normalized to the
total mass and the correlation factor (plotted as $1-r$) as a function
of the age obtained for H$\beta$ line are shown. The slopes for
H$\beta$ intensity are practically constant and with a value almost
equal to one. This behaviour is expected for hydrogen recombination
lines (the H$\alpha$ intensity shows similar results) because their
intensity is practically proportional to the $Q$(H$^0$) of the
ionizing cluster, which is directly related with the cluster mass.
This is coherent with the small range of the $L$(H$\beta$)/$Q$(H$^0$)
ratio shown in Fig. \ref{fig:LhbQH} for each age. Also, $1-r\approx0$
accounts for this proportionality.

Lines that have power-laws with $\beta\approx 1$ are particularly
interesting because, as we have said, they vary linearly with the
cluster mass, or equivalently with $Q$(H$^0$), just as the continuum
produced by the stellar cluster. In general, the use of line ratios
with lines that have similar $\beta$ values provide indices that are
independent on the cluster mass, and so, such indices can be
calibrated with observations of clusters without the need to worry
about the cluster size (if IMF sampling effects are not taken into
account). This is the case of $EW$(H$\beta$), as can be seen in Fig.
\ref{fig:EWhb}, because the intensity of the continuum is proportional
to the cluster mass.

\subsection{Collisional lines}
\label{subsec:Collines}

\begin{figure*}
\begin{center}
\includegraphics[width=16cm]{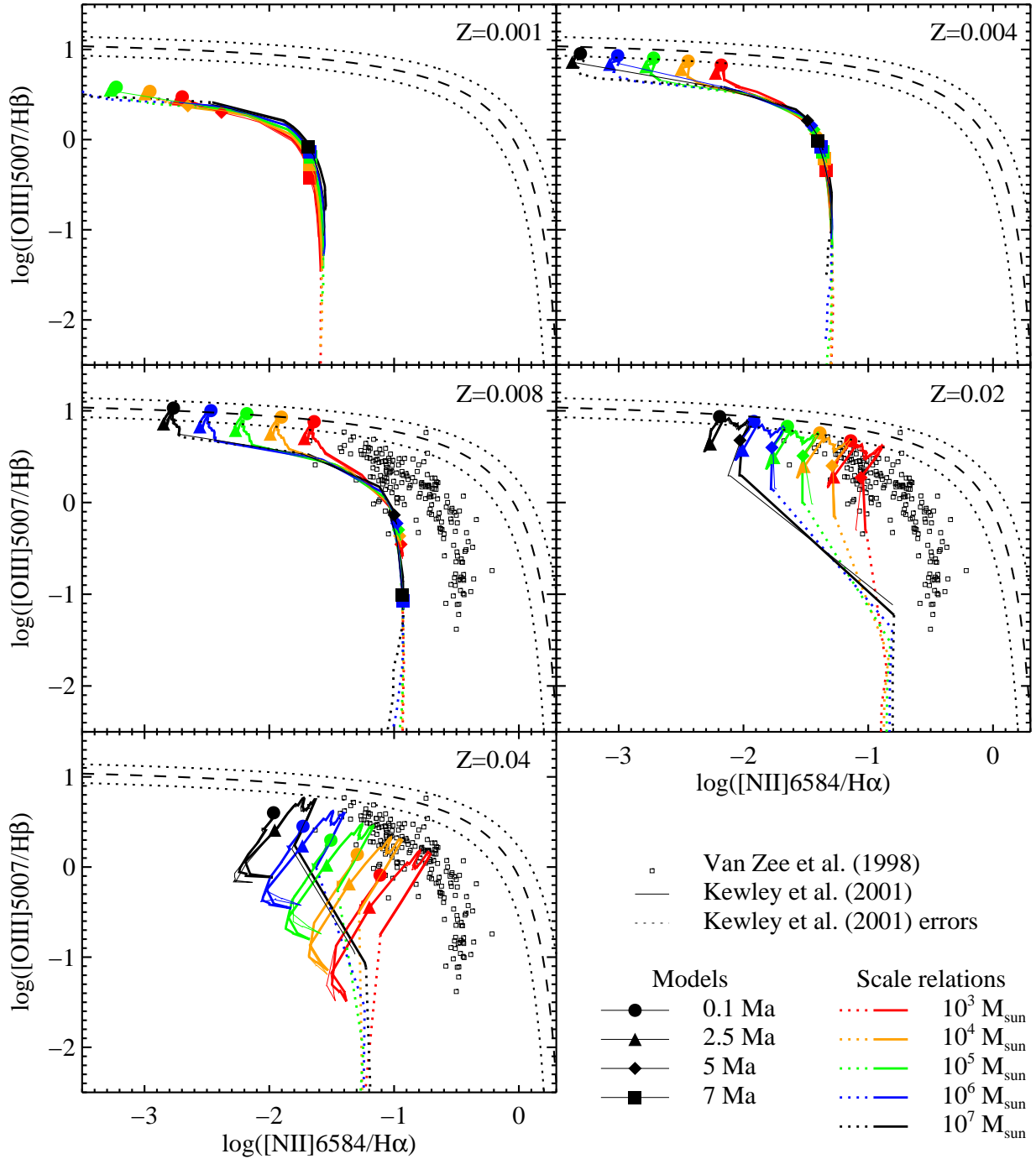}
\end{center}
\caption[]{[\ion O{iii}] 5007/H$\beta$ vs [\ion N {ii] }6584/H$\alpha$
  diagnostic diagrams for each metallicity. Grey scale lines
  represent results from models and scale-relations for different
  cluster mass. Narrow lines with symbols represent the age evolution
  of models that fulfill the $EW > 1$ \AA ~criteria; different symbols
  correspond to different ages. Dotted/solid lines are the results of
  applying the scale relations to the individual cluster masses, being
  the dotted part the region where the $EW$ (as computed by Eq.
  \ref{eq:ew}) of any of the lines involved is lower than 1 \AA. For
  reference we also show \citet{vZ98} observations (open black
  squares), and the \citet{Ketal01} curve (long dashed) with its error
  limits (short dashed curves).}
\label{fig:OIIIvsNII}
\end{figure*}

\begin{figure*}
\begin{center}
\includegraphics[width=16cm]{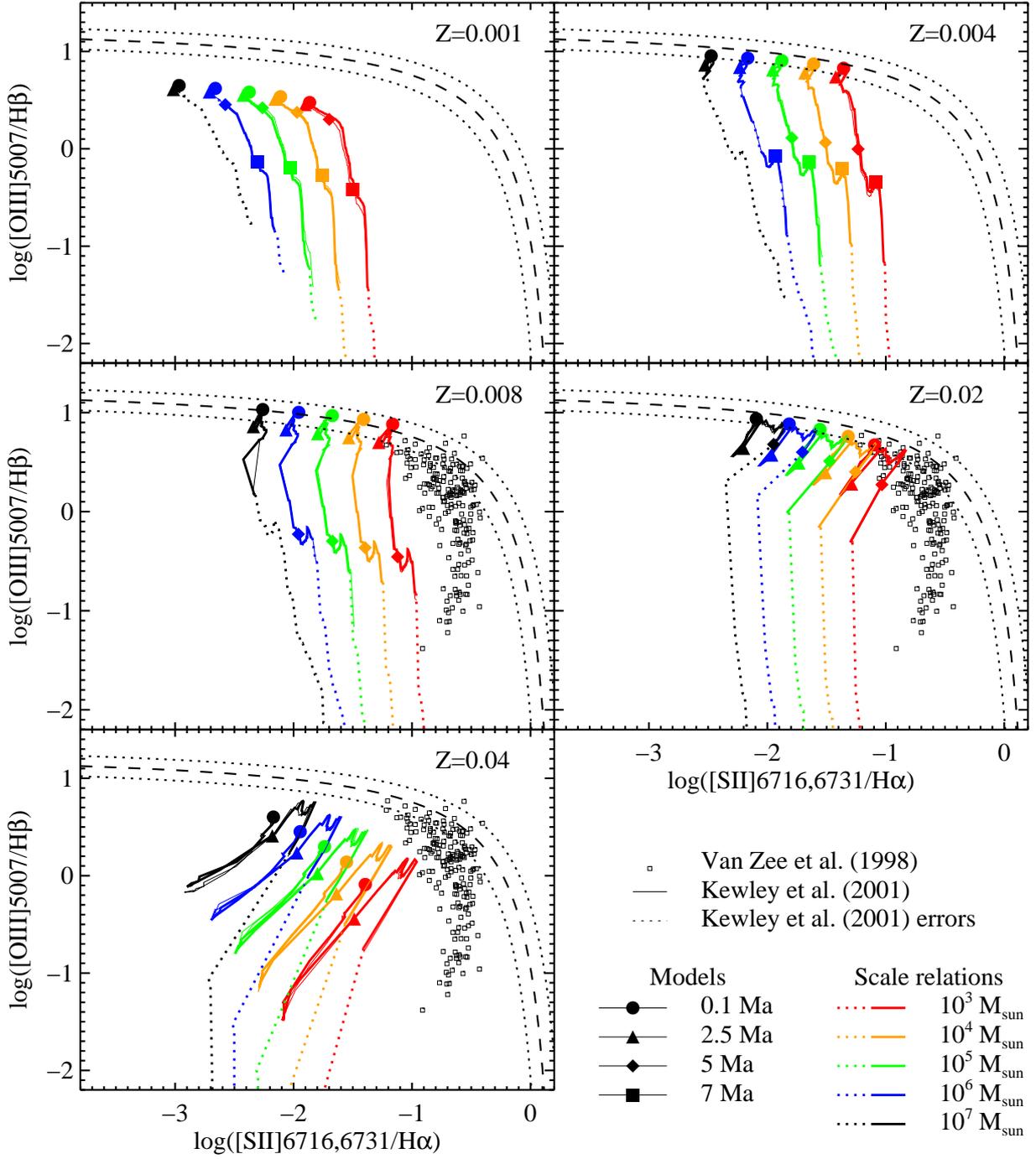}
\end{center}
\caption[]{\ion O{iii} 5007/H$\beta$ Vs \ion S{ii} 6716,6731/H$\alpha$
  diagnostic diagrams for each metallicity. Symbols as in Fig.
  \ref{fig:OIIIvsNII}.}
\label{fig:OIIIvsSII}
\end{figure*}

Following the same method described before we have obtained
scale-relations for some relevant collisional lines. In Fig.
\ref{fig:collines} the slopes for [\ion O{ii}] $\lambda$ 3727 \AA,
[\ion O{iii}] $\lambda$ 5007 \AA, [\ion N {ii}] $\lambda$ 6584 \AA,
[\ion S{ii}] $\lambda$ 6716 \AA, and [\ion S{iii}] $\lambda$ 9069 \AA
~\ are shown. The grey scale colors of the lines have the same meaning
as in Fig. \ref{fig:EWhb}. We have represented with dotted lines the
ages where the scale relations computed through Eq. \ref{eq:ew} do not
produce $EW \ge 1$ \AA ~for all masses in the $10^3 - 10^6$ M$_\odot$
mass range (we have excluded the $10^7$ M$_\odot$ case, since it is a
purely academic case with no implications for observations). For
these emission lines $r > 0.98$ in all the age range, therefore within
our models the logarithm of the intensity of the emission lines is, to
a good approximation, linear with the logarithm of the cluster mass.

The slopes of [\ion O{ii}] $\lambda$ 3727 \AA, [\ion N {ii}] $\lambda$
6584 \AA, and [\ion S{iii}] $\lambda$ 9069 \AA ~\ have a similar
behaviour. In the three cases, there are two regimes, a young one with
slopes around 0.7-0.8, and an a older one with slopes around 1
(excluding the Z=0.04 models). The transition between the two regimes
happens in the age range 4 - 5.5 Ma and it is due to a softening of
the ionizing continua in such age range. Comparing with the evolution
of the cluster $T_{\mathrm{eff}}$, defined as in \citet{MHK91}, the
change in the slope is associated with a cluster $T_{\mathrm{eff}}
\sim3.5\times10^4$ K following the results from \citet{CMH94} and
\citet{Cetal02b}\footnote{{\tt
    http://www.laeff.inta.es/users/mcs/SED/index.html}}. Such cluster
$T_{\mathrm{eff}}$ is not due to the predominance of a specific type
of star but to a mix of different types of stars.

Since the different values of the slopes indicate different
dependences with the mass, and taking into account that the continuum
is proportional to the cluster mass, it is straightforward that the
equivalent widths also present different dependences with the mass.
This can be seen in Fig. \ref{fig:EWO}a in which $EW$([\ion O{ii}]
3727) is shown for $M \geq 10^3$M$_\odot$ and all the metallicities.
There is a clear dependence of $EW$( [\ion O {ii}] $\lambda$ 3727)
with the mass for young ages (slopes different from 1). On the other
hand, the $EW$ is almost independent on the cluster mass for old ages.

[\ion O{ii}] $\lambda$ 3727 \AA ~\ and $EW$( 3727) are used as star
formation rate (SFR) indicators \citep[][ among others]{GBH89,K92,
 RGTT02}. The method is based on linear relations between [\ion
O{ii}] $\lambda$ 3727 \AA ~\ and H$\alpha$ or H$\beta$. Many authors
have studied the dependence of these relations on the luminosity
\citep{JFF01, Aetal03}, the reddening \citep{RGTT02, Aetal03} and
abundances \citep{JFF01,KGJ04, Mouhetal05, Mousetal06}. From the
value of the slopes for [\ion O{ii}] $\lambda$ 3727 \AA ~\ and
H$\alpha$ (being the last similar to the ones shown in Fig.
\ref{fig:Hb}), we obtained that the [\ion O{ii}] 3727/H$\alpha$ ratio
roughly depends on $M^{-0.3}$ in the 0.1-4.5 Ma range, having a
dependence flatter than $M^{0.05}$ for larger ages. Therefore at very
early ages the SFR estimation has a non negligible dependence on the
mass of the ionizing cluster of the observed \ion H{ii} regions.
However, it is a good SFR indicator for older ages. Our study
suggests that a recalibration of [\ion O{ii}] $\lambda$ 3727 \AA ~\
using only old (age $>$ 4.5 Ma) clusters would improve its use as an
SFR indicator and that the method should be applied to galaxies
dominated by \ion H {ii} regions older than 4.5 Ma.

It is noteworthy that the slope of [\ion O{iii}] $\lambda$ 5007 \AA ~\
is nearly 1. Considering all the metallicities except $Z$=0.04 and
ages lower than 7 Ma, we can adopt a value of 1.05 as a representative
slope for this emission line. In such case we obtain $EW$([\ion
O{iii}] 5007) $\propto M^{0.05}$. Therefore $EW$([\ion O{iii}] 5007)
seems to be a suitable age indicator as \citet{SL96} had proposed.
Nevertheless the dependence with the cluster mass is large enough to
produce a non-negligible uncertainty in the age. This can be checked
in Fig. \ref{fig:EWO}b where the equivalent width of [\ion O{iii}]
$\lambda$ 5007 \AA ~\ is shown. Comparing it with Fig. \ref{fig:EWhb},
it can be seen that the sequences for equal metallicity span a larger
$EW$-range than for $EW$(H$\beta$). Moreover there are more situations
of age degeneracy if [\ion O{iii}] $\lambda$ 5007 \AA ~\ is used.
Therefore caution is needed when using this line for age estimations.
In Fig. \ref{fig:EWO}b it can also be seen that the age range for the
application of $EW$([\ion O{iii}] 5007) as an age indicator is smaller
than for $EW$(H$\beta$), because $EW$([\ion O{iii}] 5007), and
therefore the detection of the line, decreases faster than
$EW$(H$\beta$).

The slopes of [\ion S{ii}] $\lambda$ 6716 \AA ~\ are smaller than 1
except for $Z$=0.04 at older ages. This means that the $EW$ and ratios
of [\ion S{ii}] $\lambda$ 6716 \AA ~\ with H$\alpha$ and H$\beta$ will
decrease with increasing cluster mass.

The reliability of these scale-relations can be checked comparing the
results obtained with them with the results obtained directly from the
models. The comparisons have been made with some diagnostic diagrams
and computing some relevant parameters. At the same time, clues of the
influence of the cluster mass on such diagrams and parameters have
been obtained.

\subsubsection{Diagnostic diagrams}
\label{subsubsec:diagdiag}

We have used two diagnostic diagrams to compare the scale-relations
with the models. Let us remind that scale-relations built with models
in the whole mass range (1-10$^7$ M$_{\odot}$) do not agree well with
the results obtained directly from the models, especially in the
high-mass range where the luminosity is overestimated. The inclusion
of low mass clusters, which is just an academic case, produces a
concave curvature in the $\log M - \log L$ plane, so the inclusion of
low mass clusters produces a steeper slope (and a lower normalized
luminosity). As we explain in section
\ref{subsubsection:scalerelations} we primarily seek a good agreement
in the high-mass range. For this reason our final scale-relations
have been computed only with models with $M \ge 10^3$ M$_\odot$. This
clearly improves the results of scale-relations in the high-mass
range; furthermore, we have also checked that for low masses the
results are acceptable (though overestimated). In the following, we
focus only on the simulations with clusters larger than $10^3$
M$_\odot$.

In Fig. \ref{fig:OIIIvsNII} the [\ion O{iii}] 5007/H$\beta$ vs [\ion N
{ii}] 6584/H$\alpha$ diagram proposed by \citet{BPT} is shown. Narrow
lines with symbols represent the age evolution of models that fulfill
the $EW > 1$ \AA ~criterion. Different symbols correspond to different
ages. Note that the models do not fulfill the $EW$ for all ages (and
they have not been plotted in such cases). Dotted/solid lines are the
results of applying of the estimations of the line intensities with
the scale relations obtained previously to the individual cluster
masses, the dotted part being the ages where the $EW$ of any of the
lines involved in the diagram is lower than 1 \AA. The theoretical
upper limit for starbursts from \citet{Ketal01} (black dashed line)
and its estimated errors (black dotted line) are also shown for
reference.

In general, there is a good agreement between the models and the
results obtained using the scale relations, and there are only
noticeable discrepancies for the $Z=0.04$ case. It is also interesting
to find in the $Z=0.001$ case that only the less massive clusters
produce $EW$([\ion{N}{ii}]) larger than 1. It is consistent with the
results of Fig. \ref{fig:collines}, which show that, given the slope
lower than 1 of the $L([\ion{N}{ii}])$ scale relation, the
$EW$([\ion{N}{ii}]) decreases as the cluster mass increases for young
ages. Our models and scale relations show that for fixed metallicity
and $T_{\mathrm{eff}}$, i.e. constant age, there is a sequence in
decreasing Q(H$^{0}$), or equivalently in decreasing mass. In general,
models and scale relations have trends similar to, or at least
compatible with, the observations.

Some of our scale-relations (and models) are slightly above the
\citet{Ketal01} curve. Such models correspond to clusters of 10$^{7}$
M$_{\odot}$ and metallicities in the 0.004-0.02 range. \citet{Ketal01}
estimated an error of $\pm$0.1 dex due to potential errors in the
modelling. Taking into account such errors, all our scale-relation
estimations agree with the \citet{Ketal01} limits.

In Fig. \ref{fig:OIIIvsSII} the [\ion O{iii}] 5007/H$\beta$ vs [\ion
S{ii}] 6716+6731/H$\alpha$ diagnostic diagram is shown. The results
estimated from the power-law scale-relations follow the trends of the
models, and also reproduce some of the features (peaks, turn-off and
$EW$) that the sequences of models present. The diagram shows clearly
the effects of the cluster mass in the [\ion S{ii}] 6716+6731, which
produce a serie of parallel sequences. Again, there is a good
agreement between models and scale relations, with the larger
discrepancies at $Z=0.04$. Scale relations show that it is more
difficult to detect the [\ion S{ii}] lines in more massive clusters.
Since the slope in the $\log M - \log L$ relation is almost constant
for all ages and lower than one, both the [\ion S{ii}]
6716+6731/H$\alpha$ and the $EW$ of the line decreases with increasing
cluster mass. In this diagram, observations and models/scale
relations have a poorer agreement than in the [\ion O{iii}]
5007/H$\beta$ vs [\ion N{ii}] 6584/H$\alpha$ diagram. However, notice
that [\ion S{ii}] $\lambda\lambda$ 6716, 6731 \AA, as [\ion O {i}]
$\lambda$ 6300 \AA, are likely to be affected by shocks and since
Cloudy does not include shocks, the results for those lines are not
complete.

From these comparisions we conclude that, although not perfect, the
scale-relations are suitable tools for cluster masses larger than
$10^3$ M$_\odot$. 

\subsubsection{The $\eta$ parameter, $R_{23}$ and $S_{23}$}
\label{subsubsec:parameters}

\begin{figure*}
\begin{center}
  \includegraphics[width=17cm]{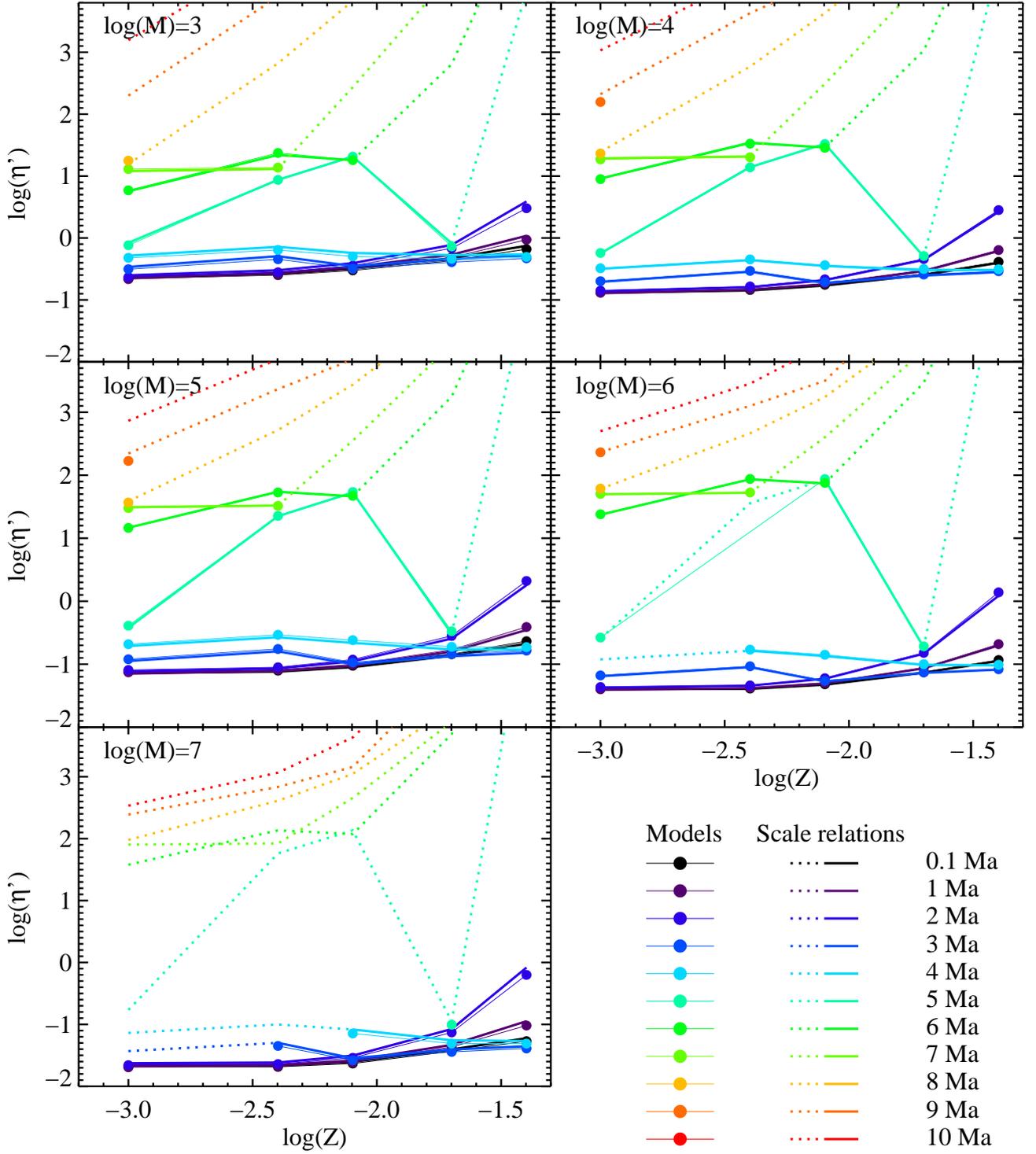}
\end{center}
\caption[]{$\eta^\prime$ parameter computed with scale-relations
  (dotted/solid broad lines) and models (narrow lines) for each
  cluster mass. Grey scale colors represent ages.}
\label{fig:eta}
\end{figure*}

The $\eta$ parameter was defined by \citet{VP88} as:

\begin{equation}
\eta ={ {\mathrm{O}^{+}/\mathrm{O}^{++}} \over {\mathrm{S}^{+}/\mathrm{S}^{++}}}\: , 
\label{eq:etadef}
\end{equation}

\noindent and it is related with the observational ratio:

\begin{equation}
\eta^\prime = {{{[\ion O{ii}]\; 3727+3729} / {[\ion O{iii}]\; 4959+5007}} \over {{[\ion S{ii}]\; 6717+6731} / {[\ion S{iii}]\; 9069+9532}}}.
\label{eq:etaprimedef}
\end{equation}

The slopes of the scale-relations computed for each of the lines
involved in the ratio can be used to estimate the $\eta^\prime$
dependence on mass. As can be seen in Fig. \ref{fig:collines}, for
each line the computed slopes have similar values and trends for all
metallicities except for $Z$=0.04. Thus, as a rough approximation, we
can define for each line a representative slope and obtain the
following proportionalities valid for all the metallicities considered
except for $Z$=0.04:

\begin{eqnarray}
\label{eq:prop1}
I([\ion O{ii}] 3727) & \propto & \left\{ 
\begin{array}{l l}
M^{0.75} \quad \mathrm{if} \quad  \mathrm{age}<4.5\: \mathrm{Ma}, \\
M^{1.03} \quad \mathrm{if} \quad \mathrm{age}>4.5\: \mathrm{Ma}; \\
\end{array} \right. \\
\label{eq:prop2}
I([\ion O{iii}] 5007) & \propto & M^{1.05};  \\
\label{eq:prop3}
I([\ion S{ii}] 6716)  & \propto &  M^{0.73};  \\
\label{eq:prop4}
I([\ion S{iii}] 9069) & \propto  & \left\{ 
\begin{array}{l l}
M^{0.77} \quad \mathrm{if} \quad  \mathrm{age}<4.5\: \mathrm{Ma}, \\
M^{1.03}  \quad \mathrm{if} \quad \mathrm{age}>4.5\: \mathrm{Ma}. \\
\end{array} \right. 		
\end{eqnarray}

[\ion O{iii}] $\lambda$ 4959 \AA, [\ion S{ii}] $\lambda$ 6731 \AA ~\
and [\ion S{iii}] $\lambda$ 9532 \AA ~\ present dependences with the
cluster mass similar to their associated lines from above. Using these
proportionalities we obtain for $\eta^\prime$:
 
\begin{equation}
\eta^\prime\propto \left\{
\begin{array}{l l}
M^{-0.26} \quad \mathrm{if} \quad  \mathrm{age}<4.5\: \mathrm{Ma}. \\
M^{0.28}  \quad \mathrm{if} \quad \mathrm{age}>4.5\: \mathrm{Ma}. \\
\end{array} \right.
\end{equation}

These trends can be seen in Fig. \ref{fig:eta} where $\eta^\prime$
versus log($Z$) is represented for each cluster mass. The grey scale
colors, unlike in previous figures, indicate the age. Results
obtained directly from models are shown with circles connected by
narrow lines. We have only plotted the models that fulfil the $EW$
criteria. Results computed with scale-relations are represented with
dotted/solid lines with a larger line weight. Dotted parts of the
lines connect metallicity points where either point does not fulfill
the $EW$ criteria as obtained by the scale laws. The agrement within
the models and scale relations is remarkable. At ages lower than 4.5
Ma, scale-relations produce $\eta\prime$ values that decrease with
increasing mass, while at larger ages $\eta\prime$ increases with
mass. This behaviour is due to the variation of the scale relation of
the [\ion{S}{iii}] and [\ion{O}{ii}] lines.

\begin{figure*}
  \begin{center}
  \includegraphics[width=17cm]{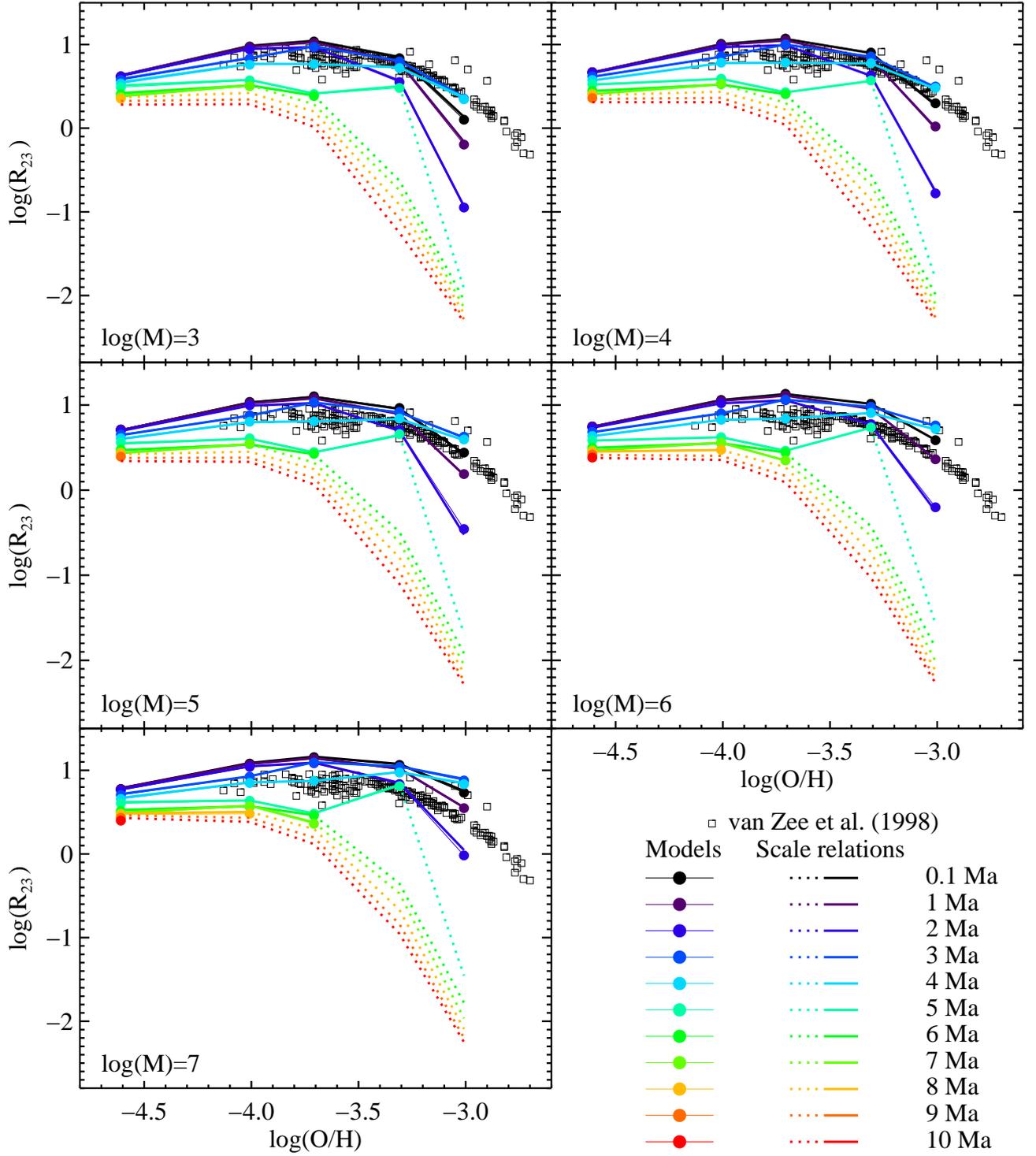}
  \end{center}
  \caption[]{$R_{23}$ for each cluster mass. Black squares:
    \citet{vZ98} observations. Rest of symbols as in Fig.
    \ref{fig:eta}.}
\label{fig:r23}
\end{figure*}

The $R_{23}$ parameter was proposed as an abundance calibrator by
\citet{Petal79}, and is defined as:

\begin{equation}
R_{23} = {{{[\ion O{ii}]\; 3727} + {[\ion O{iii}]\;4959+5007}} \over {\mathrm{H}{\beta}}}.
\label{eq:R23def}
\end{equation}

Following the same method as for $\eta^\prime$, we estimate the
dependence of $R_{23}$ with the cluster mass. Taking into account that
H$\beta\,\propto\,M$ and using Eqns. \ref{eq:prop1} and
\ref{eq:prop2}, we obtain:

\begin{equation}
R_{23} \propto \left\{
\begin{array}{l l}
A_{R23}M^{-0.25}+B_{R23}M^{0.05} \quad \mathrm{if} \quad  \mathrm{age}<4.5\: \mathrm{Ma}. \\
A_{R23}M^{0.03}+B_{R23}M^{0.05}  \quad \mathrm{if} \quad \mathrm{age}>4.5\: \mathrm{Ma}. \\
\end{array} \right. 
\end{equation}

\noindent where the A$_{R23}$ and B$_{R23}$ factors are obtained from
the intensity of the lines used in $R_{23}$ normalized to the total
mass, therefore they depend on age and metallicity.

Recalling that the continuum has a slope equal to 1 (see sec.
\ref{subsubsection:scalerelations}) the same result is obtained if the
$EW$ of the lines are used as \citet{KP03} proposed.

In Fig. \ref{fig:r23} the values obtained for $R_{23}$ are shown. The
dependence in mass estimated above, in spite of being small, can be
appreciated when comparing carefully the plots for different masses.
The agreement between scale-relations and models is better than for
the $\eta^\prime$ parameter.

\begin{figure*}
  \begin{center}
  \includegraphics[width=17cm]{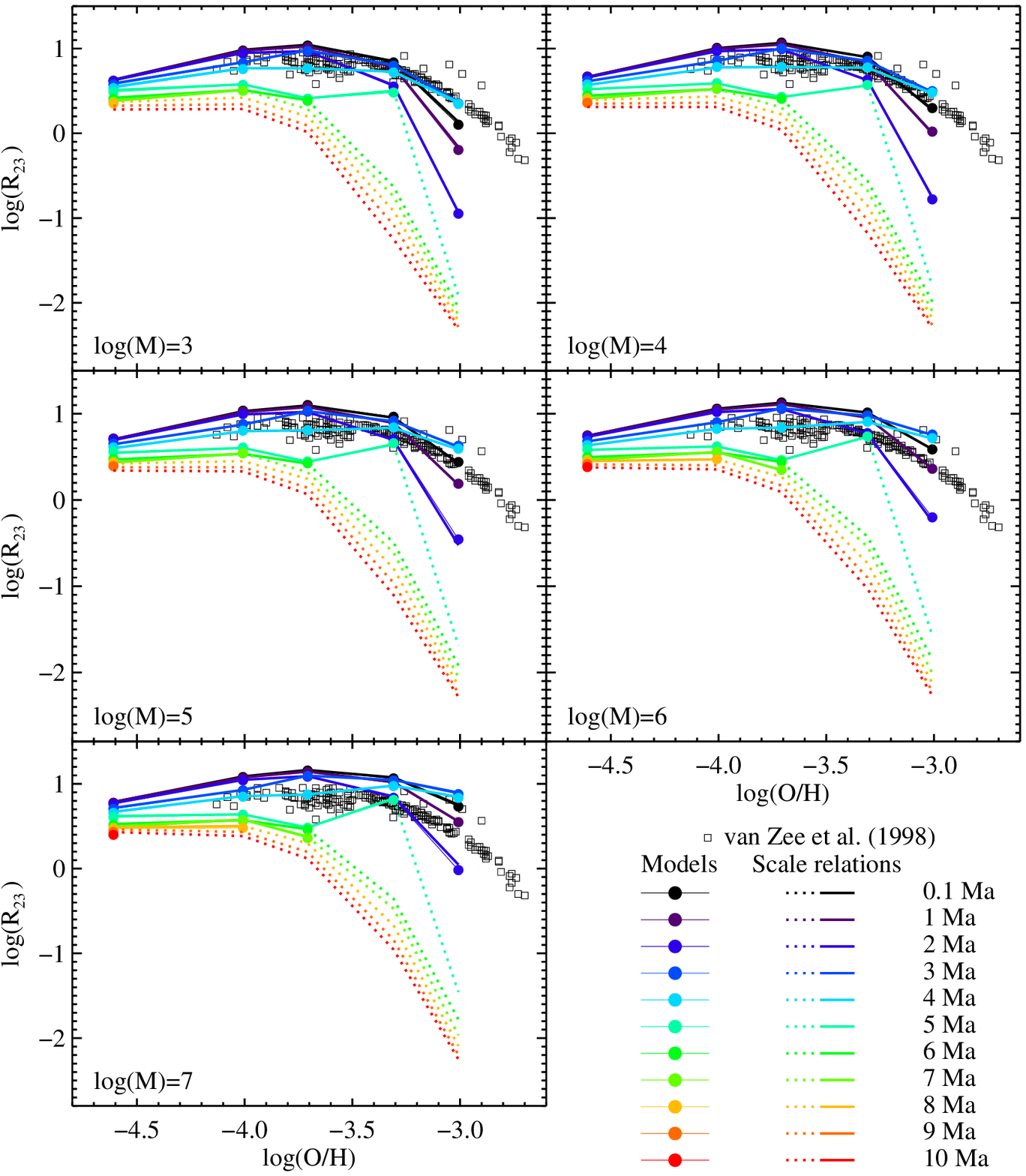}
\end{center}
\caption[]{$S_{23}$ for each cluster mass. Symbols as in Fig. \ref{fig:eta}}
\label{fig:s23}
\end{figure*}

The $S_{23}$ parameter was defined by \citet{VE96} in a way similar to
$R_{23}$:

\begin{equation}
S_{23} = {{{[\ion S{ii}] 6717+6731} + {[\ion S{iii}] 9069+9532}} \over {\mathrm{H}{\beta}}}.
\label{eq:S23def}
\end{equation}

For this parameter we obtain a mass dependence of the form:

\begin{equation}
S_{23} \propto \left\{
\begin{array}{l l}
A_{S23}M^{-0.27}+B_{S23}M^{-0.23} \quad \mathrm{if} \quad  \mathrm{age}<4.5\: \mathrm{Ma}. \\
A_{S23}M^{-0.27}+B_{S23}M^{0.03}  \quad \mathrm{if} \quad \mathrm{age}>4.5\: \mathrm{Ma}. \\
\end{array} \right .
\end{equation}

\noindent where the A$_{S23}$ and B$_{S23}$ factors depend on age and
metallicity.

As $R_{23}$, $S_{23}$ also shows a dispersion of the results of the
models due to different cluster masses (see Fig. \ref{fig:s23}). A
relevant fact is the variation of $S_{23}$ with the age through the
mass range. As the mass increase the sequences for older ages rise and
the lower ages ones go down. This behaviour can be explained with the
different dependences of $S_{23}$ with the mass for ages below and
over 4.5 Ma.

\section{Summary of results and discussion}
\label{sec:discussion}

Using models of \ion H{ii} region ionized by clusters of different
masses, we have studied the variation of the \ion H{ii} region
emission with the ionizing cluster mass. For this task,
scale-relations in form of power-laws between the intensities of some
of the most relevant emission lines and the ionizing cluster mass have
been obtained. The reliability of these power-law scale-relations has
been checked by means of diagnostic diagrams. We check that the
general agreement between power-law scale-relations and models is
better if we compute the power-laws using only the models with $M \geq
10^3$ M$_\odot$, with the intensities of collisional lines being
overestimated by the power law relations in the case of clusters with
masses lower than $10^3$ M$_\odot$. However, the ionizing flux of such
low mass clusters cannot be described by the average ionizing flux
obtained by synthesis models (c.f. Paper I). 

We want to stress that, due to the assumptions in the modelling and
the simple linear approximation, these scale-relations are just a
partial first approach from which we can obtain an overall view of the
dependence of line intensities with the cluster mass, but these scale
relations must not be used as an {\it exact} representation of the
variation of the intensities of the emission lines with the cluster
mass. In this sense, our scale-relations, although incomplete, are
very useful tools that describe {\it mean} luminosity functions.
Therefore they could be used when considering the mass dependence of
the properties of massive \ion H {ii} regions samples, but should not
be applied to individual objects. An example of the limitations of
our scale-relations is that they cannot predict the variation of the
H$\alpha$/H$\beta$ variation of the models with respect to the
normally assumed value, variation that induce a spurious extinction
effect.

In spite of these limitations, simply estimating the trend of the
luminosity distribution function allows to make interesting
predictions. The exponents of the power-law scale-relations indicate
us the mass dependence of the line luminosity. With this dependence it
is possible to estimate the influence of the cluster mass on $EW$s and
on some relevant line intensities ratios. If a quantity has no
dependence on cluster mass it can be calibrated with observations with
no need of age and cluster mass estimations, otherwise a correction in
mass should be applied. We check that $EW$(H$\beta$) is almost
insensitive to mass, while $EW$([\ion O{iii}] 5007) has a small
dependence. This implies that $EW$([\ion O{iii}] 5007) is not as good
as an age indicator as $EW$(H$\beta$), and that we have to take into
account the cluster mass when estimating ages with $EW$([\ion O{iii}]
5007). We also explore the variation of $EW$([\ion O{ii}] 3727) with
mass cluster and we obtain an important mass dependence for ages lower
than $\sim$4.5 Ma, depending on metallicity, and almost no mass
dependence at older ages. The different mass dependences of [\ion
O{ii}] $\lambda$ 3727 \AA, [\ion O{iii}] $\lambda$ 5007 \AA , and
therefore of their $EW$s, offer a possible explanation for the
different dispersions in their respective log($EW$/$EW$(H$\beta$)) vs
log(I/I(H$\beta$)) plots \citep{KP03}. The mass dependence of the
[\ion O{ii}] 3727/H$\alpha$ ratio, similar to $EW$([\ion O{ii}] 3727),
implies that the [\ion O{ii}] $\lambda$ 3727 \AA ~\ intensity is not a
good SFR indicator if clusters with ages lower than 4.5 Ma are
present. This suggests that the [\ion O{ii}] $\lambda$ 3727 \AA ~\
intensity should be recalibrated and that its use should be limited to
galaxies dominated by \ion H{ii} regions older than 4.5 Ma.

\begin{figure}
  \resizebox{\hsize}{!}{\includegraphics[width=7cm]{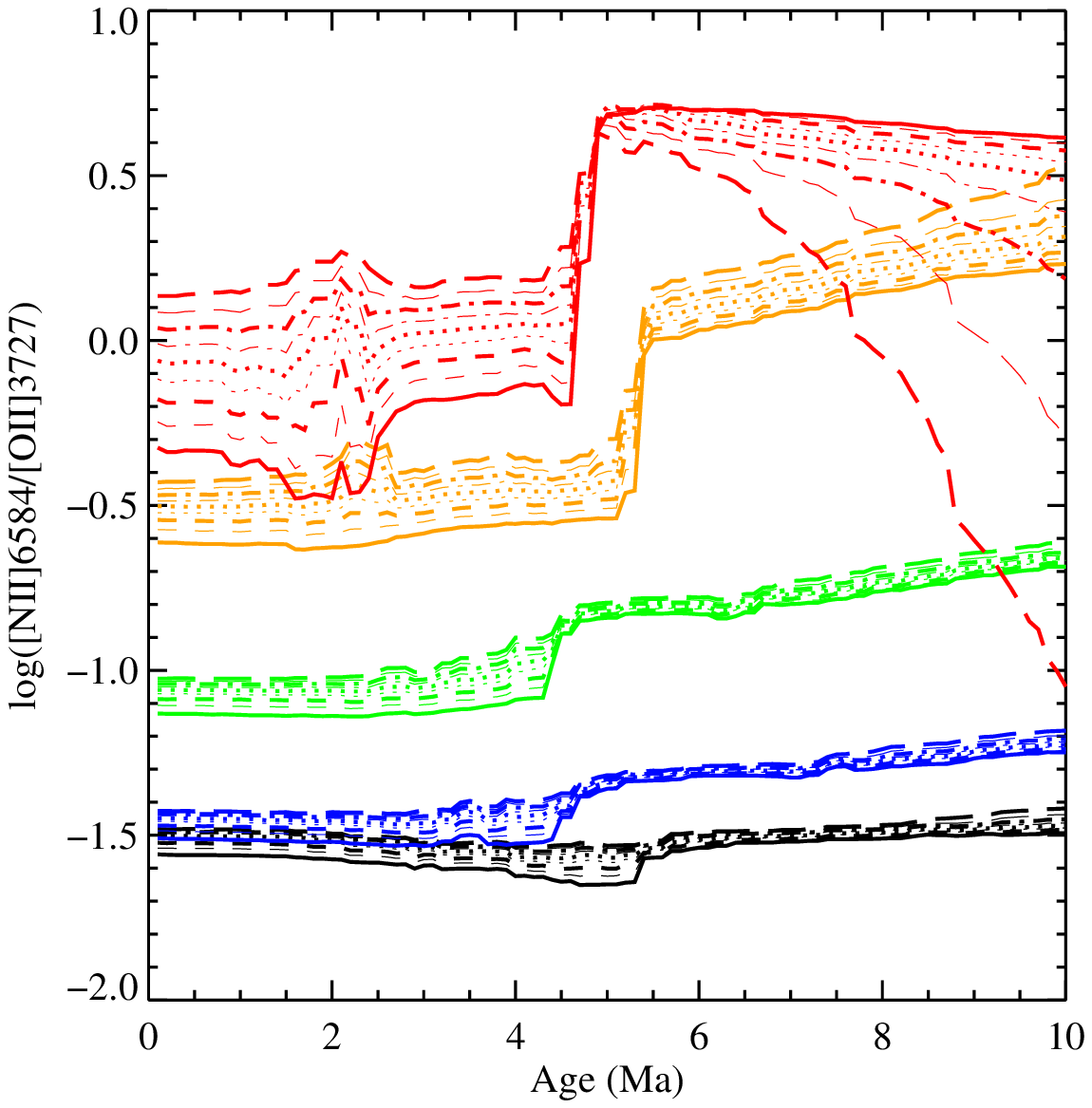}}
  \caption[]{\ The [\ion N {ii}] 6584/[\ion O{ii}] 3727 ratio versus
    age for clusters with $M \geq 10^3$. Symbols as in Fig.
    \ref{fig:EWhb}.}
\label{fig:OIINIIratio}
\end{figure}

The power-law scale-relations also allow us to identify emission line
ratios with small mass dependence. This is the case of [\ion O{ii}]
3727/[\ion N {ii}] 6584 ratio, proposed as an abundance indicator by
\citet{vZ98} and \citet{Dop00}. The values and behaviour of the
slopes for these two lines are similar, with greater differences at
the highest metallicities. Therefore their ratio will have a small
mass dependence at any age. This can be seen in Fig.
\ref{fig:OIINIIratio}, where the [\ion N {ii}] 6584/[\ion O{ii}] 3727
ratio is plotted versus the age. The thin sequences obtained for low
metallicities account for the small mass dependence. This clearly
helps to separate the sequences for differents metallicities, allowing
a better abundance estimation. The drop of [\ion N {ii}] 6584/[\ion
O{ii}] 3727 for Z=0.040 (long-dashed red line) at ages older than 6 Ma
is due to a higher contribution of recombination emission to the [\ion
O{ii}] $\lambda$ 3727 \AA ~\ luminosity.

The scale-relations have been used together with the definition of
some parameters ($\eta\prime$, $R_{23}$ and $S_{23}$) in order to
estimate their dependence with the cluster mass. For $\eta^\prime$ and
$S_{23}$ the models show variations of these ratios with the cluster
mass that, in spite of the strong assumptions in the modelling, stress
the need of taking into account the mass of the ionizing clusters when
estimating physical properties from these ratios. For $M \geq 10^3$
M$_\odot$, $\eta^\prime$ and $S_{23}$ have opposite mass dependences
at ages earlier and older than 4.5 Ma. On the other hand, $R_{23}$
shows almost no dependence on mass. In this sense, $R_{23}$ seems a
quite robust indicator. Moreover, it also means that $R_{23}$ could be
calibrated without having to worry about the mass of the ionizing
clusters, and therefore objects of different masses can be used
together in its calibration.

Additionally, the scale relations allow us to estimate when an
emission line can be detected taking into account the continuum level.
It turns out that it is easier to detect emission lines in low mass
clusters when the scale relations have slopes lower than one. 

It can be argued that the variation with the cluster mass can also be
described by variations of other parameters, but the use of the mass
as a parameter allow to use directly the synthesis models and an
easier approach to other properties parameterized with the mass (as
the star formation history) and to the the IMF sampling problem.
Moreover, implementing the mass dependence in the calibration of an
estimation method would improve the estimations. Since including the
cluster mass in \ion H {ii} region models is a non-trivial task, and
it is not a sufficiently studied issue, the theoretical calibrations
of estimation methods with mass dependent indicators should be used
with extra caution. Therefore empirical or semi-empirical calibrations
would work better than theoretical ones since the former consider in a
natural way the variation in the cluster mass, although it is
necessary to estimate the mass by alternative methods (e.g. star
counts or dynamic effects).

As we have said, our models and scale-relations are less realistic for
the lowest masses ($M \leq 10^3$ M$_\odot$). To improve the
reliability in general, and in the low mass range in particular, an
estimation of the influence of sampling effects of the stellar IMF is
needed \citep{CMH94, Cetal00, Cetal02b, Cetal03, CL06}. This would
allow to describe properly the possible distribution of the cluster
ionizing continua. Such distribution for low mass clusters has a
large dispersion and bimodality in $T_{\mathrm{eff}}$ \citep{VCL09}.
That imply that there could be low mass clusters with an ionizing flux
equal or greater than a larger mass cluster. The effects of the
dispersion of $T_{\mathrm{eff}}$ for low mass clusters on the \ion
H{ii} region spectra are treated on \citet{VCL09} to which the reader
is referred.

\begin{figure}
  \resizebox{\hsize}{!}{\includegraphics[width=7cm]{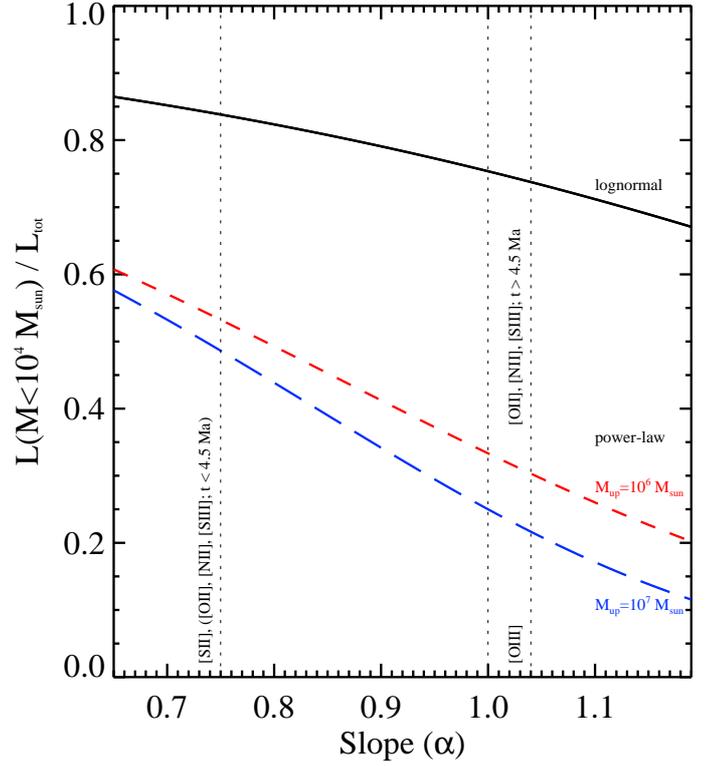}}
  \caption[]{Contribution of clusters less massive than $10^4$
    M$_\odot$ to the total luminosity as a function of the slope of
    the mass luminosity relation for different ICMF recipes.}
\label{fig:ICMF}
\end{figure}

We can obtain the same back-of-the-envelope estimation of the
influence of low mass clusters on the emission line spectra of a
non-active galaxy, and obtain a first order check of the statement by
\citet{Detal06} on the lack of influence of low mass clusters (and
hence, sampling effects) on the total line flux of collisional lines.
\cite{Detal06b} assume a lognormal ICMF

\begin{equation}
P(M) = \frac{1}{M \sigma \sqrt{2\pi}} e^{(-\frac{(\ln M - \mu)^2}{2 \sigma^2})},
\label{eq:lognormal}
\end{equation}

\noindent with $\sigma=1.7$ and $\mu = \ln 100$ M$_\odot$. Such a
distribution has a mean about 424 M$_\odot$, a median equal to 100
M$_\odot$ and a mode (maximum value) about 5 M$_\odot$. However, in
\cite{Detal06b} they argued than they only make use of this
distribution for cluster masses larger than $10^3$ M$_\odot$ and that
massive cluster wash out the stochastic effects.

We test their statement by computing the fraction of the total
luminosity of the emission line that is produced by clusters with mass
less than $M_x = 10^4$ M$_\odot$. The use of $10^4$ M$_\odot$ instead
of $10^3$ M$_\odot$ is because for $10^3$ M$_\odot$ clusters,
synthesis models includes stars more luminous than the cluster itself
\cite[the Lowest Luminosity Limit, see][~for details]{Cetal03,CL04},
so at such masses there is not only a sampling effect problem, but a
physical one in the modeling. Clusters in the $10^3 - 10^4$ M$_\odot$
are still affected by sampling effects with luminosity distributions
that are not gaussian \cite[c.f.][]{CL06}, so we take $10^4$ M$_\odot$
as a conservative lower limit for clusters affected by sampling. The
integration of a lognormal ICMF with a power-law mass dependent
luminosity is:

\begin{equation}
\int A M^\alpha P(M)\, \mathrm{d}M = - \frac{A}{2} e^{\alpha \mu +\frac{\alpha^2 \sigma^2}{2}} \,\mathrm{Erf}\left[\frac{\mu +\alpha \sigma^2 - \ln M}{\sqrt{2} \sigma}\right].
\label{eq:lognormal_int}
\end{equation}

\noindent with $\mathrm{Erf(x)}$ the gaussian error function. Hence,
the contribution of clusters less massive than $M_x$ as a function of
the slope of the mass-luminosity relation is:

\begin{equation}
\frac{L(M < M_x)}{L_\mathrm{tot}}= \frac{
\mathrm{Erf}\left[\frac{\mu +\alpha \sigma^2 - \ln M_\mathrm{low}}{\sqrt{2} \sigma}\right] - 
\mathrm{Erf}\left[\frac{\mu +\alpha \sigma^2 - \ln M_\mathrm{x}}{\sqrt{2} \sigma}\right]}
{\mathrm{Erf}\left[\frac{\mu +\alpha \sigma^2 - \ln M_\mathrm{low}}{\sqrt{2} \sigma}\right] - \mathrm{Erf}\left[\frac{\mu +\alpha \sigma^2 - \ln M_\mathrm{up}}{\sqrt{2} \sigma}\right]}.
\label{eq:lognormal_frac}
\end{equation}

We have also computed the same quantity for the case of a power-law
ICMF,

\begin{equation}
P(M) = C_\mathrm{cl} M^{-\alpha_\mathrm{cl}},
\label{eq:exp}
\end{equation}

\noindent which produces the following contribution of clusters less
massive than $M_x$ as a function of the slope of the mass-luminosity
relation:

\begin{equation}
\frac{L(M < M_x)}{L_\mathrm{tot}} = \frac{M_\mathrm{low}^{1+\alpha - \alpha_\mathrm{cl}} - M_x^{1+\alpha - \alpha_\mathrm{cl}}}
{M_\mathrm{low}^{1+\alpha - \alpha_\mathrm{cl}} - M_\mathrm{up}^{1+\alpha - \alpha_\mathrm{cl}}}.
\label{eq:powlaw_frac}
\end{equation}

Figure \ref{fig:ICMF} shows the variation of the contribution of
clusters less massive than $M_x$ to the total luminosity as a function
of the slope $\alpha$ of the mass-luminosity relation. The
approximative values of $\alpha$ for different lines has been drawn as
vertical lines. We have also assumed a $M_\mathrm{low}=10^3$ M$_\odot$
and a $M_\mathrm{up}=10^6, 10^7$ M$_\odot$ as limits of the ICMF. In
the case of a lognormal ICMF as proposed by \cite{Detal06}, the figure
shows that clusters in the $10^3 - 10^4$ M$_\odot$ range are
responsible for more than 70\% of the luminosity of the [\ion{O}{iii}]
emission lines for all ages (and [\ion{O}{ii}], [\ion{S}{iii}] and
[\ion{N}{ii}] at ages larger than 4.5 Ma). This contribution is larger
than 80\% for [\ion{S}{ii}] in all the age range (and [\ion{O}{ii}],
[\ion{S}{iii}] and [\ion{N}{ii}] at ages lower than 4.5 Ma). This
result is independent of the choice of the $M_\mathrm{up}$ : the high
mass tail of the lognormal distribution is steeper than the one of
power law distributions, so high mass clusters have a low impact on
the total flux, in opposition to \cite{Detal06b} assumption. In this
situation, the statement by \cite{Detal06b} is clearly wrong.

We have also show the case of a ICMF described by a power law with
exponent $\alpha_\mathrm{cl}=2$. In this situation, the contribution
of massive clusters to the total luminosity is more important with a
significant dependence of the choice of $M_\mathrm{up}$ in the ICMF.
However, the contribution of low mass clusters to some of the emission
lines ([\ion{O}{ii}], [\ion{S}{iii}] and [\ion{N}{ii}] at ages smaller
than 4.5 Ma) is around 50\%, and only around 20\% for [\ion{O}{iii}]
for all ages and [\ion{O}{ii}], [\ion{S}{iii}] and [\ion{N}{ii}] at
ages larger than 4.5 Ma.

Although we have not taken into account either the star formation
history or clusters less massive than $10^3$ M$_\odot$ in this
calculations, it is enough to prove that the statement about the
washing out of the stochastic effects of low mass cluster by
\cite{Detal06} needs to be revised in detail. This subject will be
addressed in another work.

\begin{figure*}
  \resizebox{\hsize}{!}{\includegraphics[width=7cm]{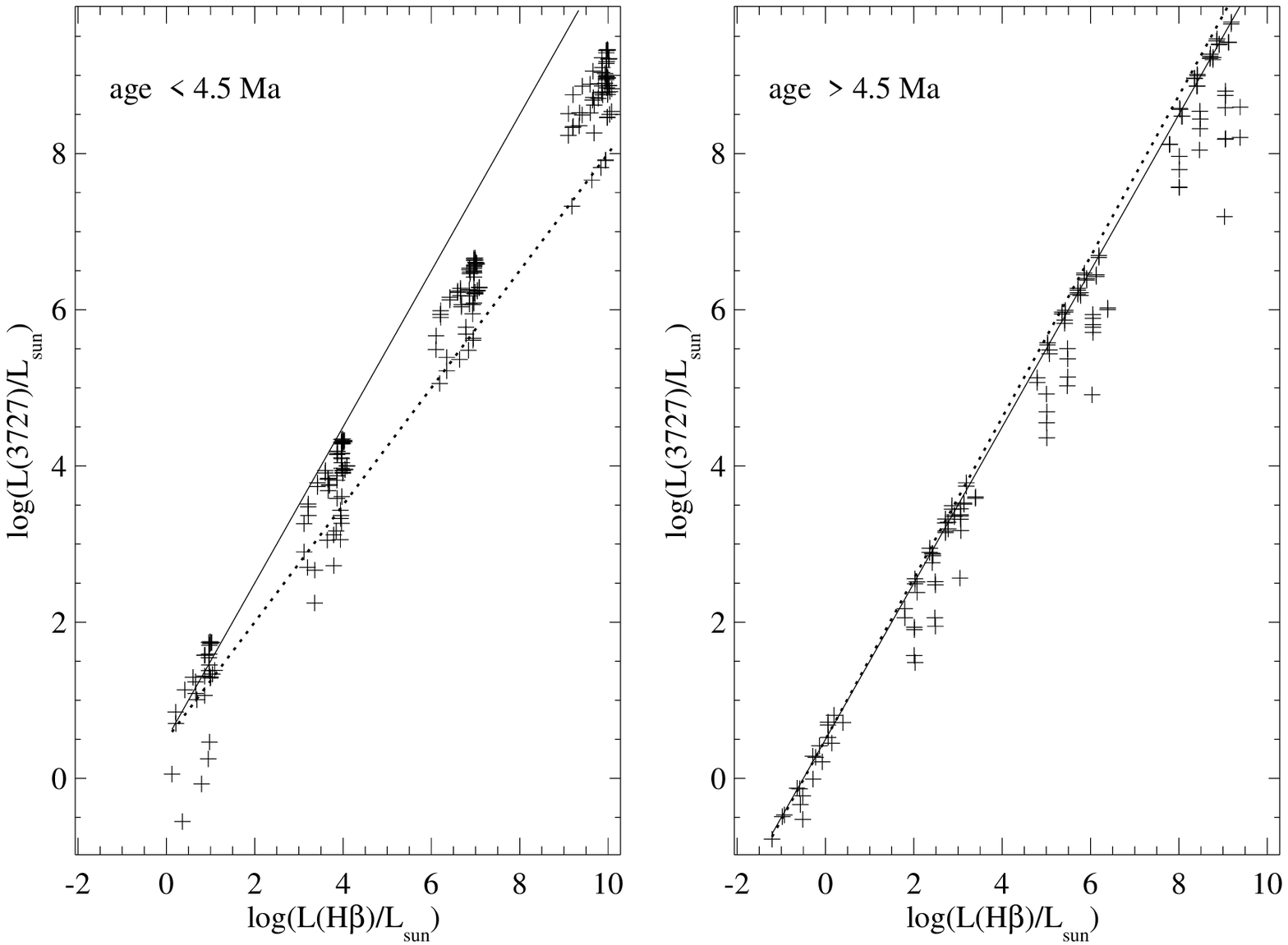}}
  \caption[]{[\ion O{ii}] $\lambda$ 3727 \AA ~/ luminosity with
    respect to the H$\beta$ luminosity for \citet{SSL01} models with
    ages lower (left) and higher (right) than 4.5 Ma. A solid line
    with slope equal to 1 and dotted lines with representative slopes
    are shown as references.}
\label{fig:SSL3727}
\end{figure*}

\begin{figure*}
  \resizebox{\hsize}{!}{\includegraphics[width=7cm]{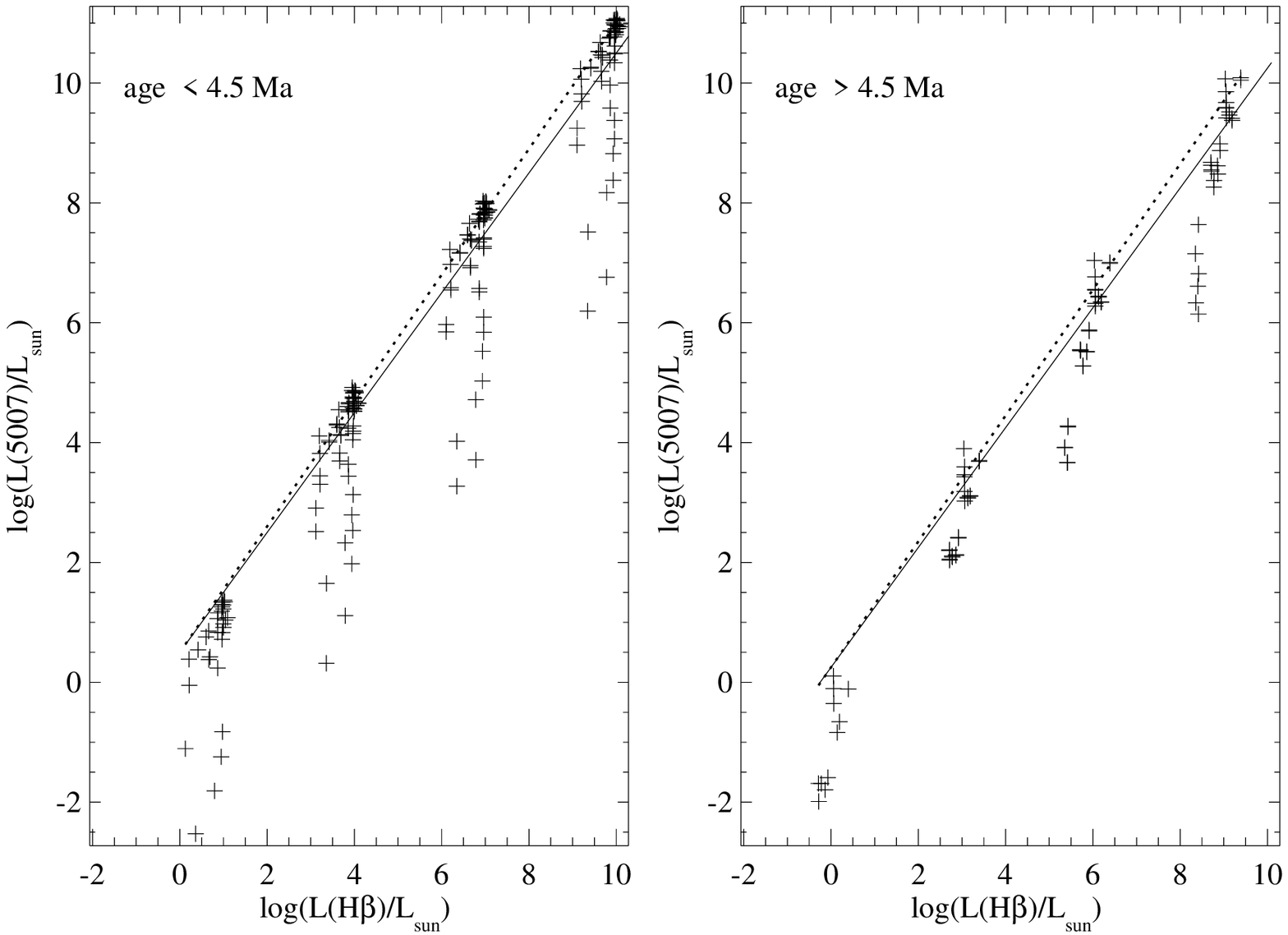}}
  \caption[]{[\ion O{iii}] $\lambda$ 5007 \AA ~/ luminosity with
    respect to the H$\beta$ luminosity for \citet{SSL01} models with
    ages lower (left) and higher (right) than 4.5 Ma. A solid line
    with slope equal to 1 and dotted lines with representative slopes
    are shown as references.}
\label{fig:SSL5007}
\end{figure*}

\begin{figure*}
  \resizebox{\hsize}{!}{\includegraphics[width=7cm]{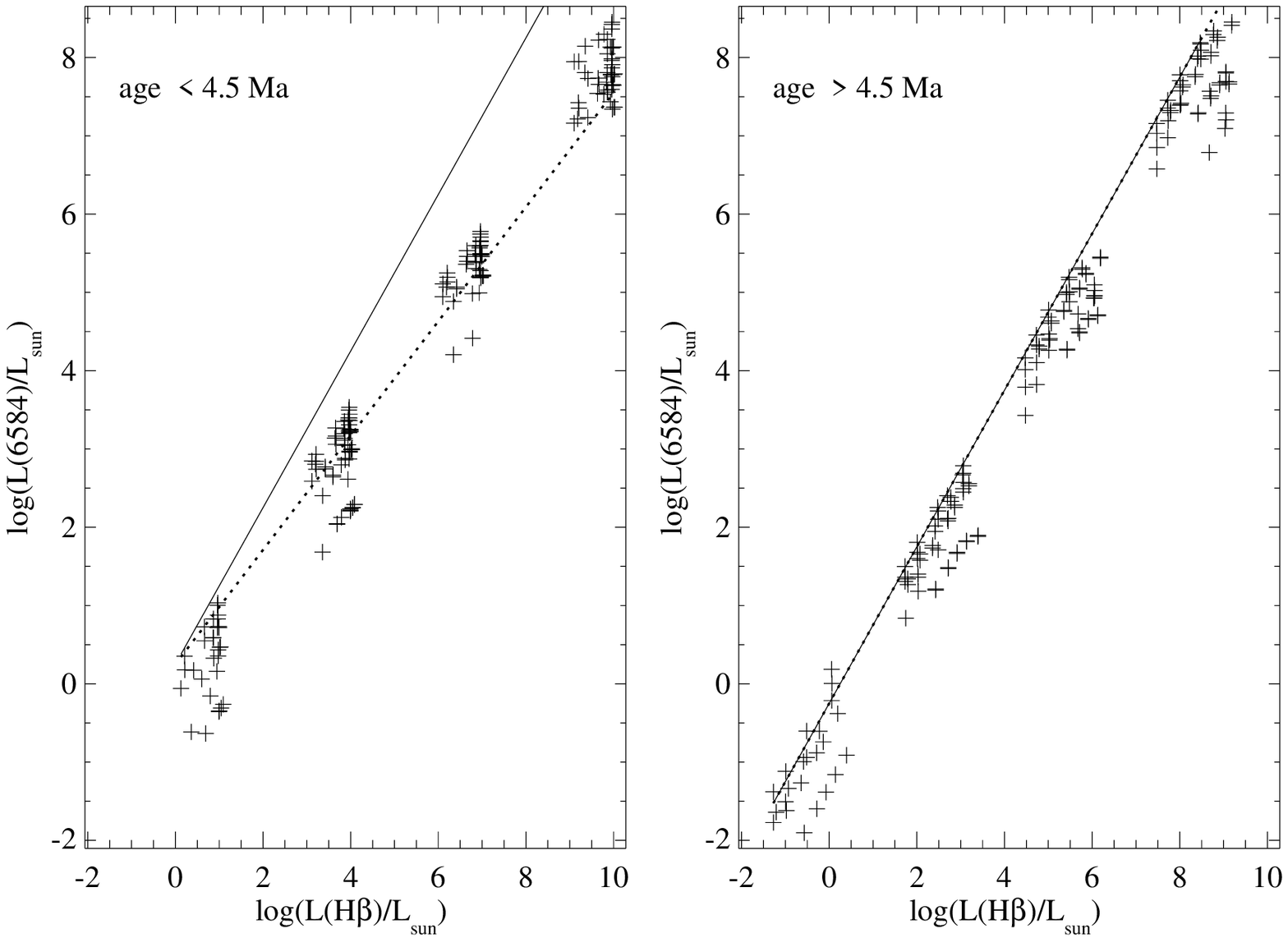}}
  \caption[]{[\ion N{ii}] $\lambda$ 6584 \AA ~/ luminosity with
    respect to the H$\beta$ luminosity for \citet{SSL01} models with
    ages lower (left) and higher (right) than 4.5 Ma. A solid line
    with slope equal to 1 and dotted lines with representative slopes
    are shown as references.}
\label{fig:SSL6584}
\end{figure*}

Finally, we have compared our scale-relations with the \citet{SSL01}
models. Those models considered several different values for density
and filling factors, moreover the inner radius varies from model to
model. For a consistence check we have only considered instantaneous
burst models with IMF upper stellar mass similar to ours. In
Figs.\ref{fig:SSL3727}, \ref{fig:SSL5007} and \ref{fig:SSL6584} the
intensities of some of the most relevant emission lines versus the
H$\beta$ intensity are shown. Since the intensity of H$\beta$ is
proportional to the mass, the tendencies shown in these plots can be
directly compared to the slopes of our scale-relations. Only the
models that have an age, mass and metallicity for which our
scale-relations predict $EW \geq$1 are plotted. Models with ages lower
and greater than 4.5 Ma are plotted separately, because around that
age many lines show a sudden change in their scale-relations slopes.
Lines with slope equal to 1 and representative slopes from the
scale-relations are shown as a reference for each emission line and
age interval. The zero-point of these reference lines has been
adjusted for a better comparison with models. For [\ion O{ii}]
$\lambda$ 3727 \AA ~\ and ages lower than 4.5 Ma the models follow a
tendency with slope smaller than 1, while for older ages the slope is
near 1. These tendencies coincide with the ones obtained with our
scale-relations (see fig.\ref{fig:collines}). The same occurs for
[\ion O{iii}] $\lambda$ 5007 \AA ~\ and [\ion N {ii}] $\lambda$ 6584
\AA ~\ (Figs. \ref{fig:SSL5007} and \ref{fig:SSL6584}). Moreover, the
scatter in the plots, which is due to the different ages,
metallicities and physical parameters of the models, corresponds to
the spread of values of the $\alpha$ coefficient of the
scale-relations. These similar tendencies indicate that our
scale-relations are robust against changes of density, inner radius
and filling factor.

\section{Conclusions}
\label{sec:conclusion}

In this work we have studied the variation of the \ion H{ii} regions
spectrum with the ionizing cluster mass. For this task we have done
photoionization models for several metallicities and ages in a cluster
mass range from $10^3$ to 10$^7$ M$_\odot$. From these models we
obtain power-law scale-relations between the emission line intensities
and the ionizing cluster mass.

These {power-law} scale-relations are very useful tools since they
allow to estimate in a simple way mass dependences of $EW$ and line
intensities ratios. Using {such} scale-relations we have checked that
$EW$(H$\beta$) is practically independent of mass, a result which
supports its use as an age {indicator}. On the other hand, $EW$([\ion
O{iii}] 5007) has a mass dependence that, in spite of being small,
implies that some caution is needed when using it as an age
{indicator}. The mass dependence obtained for [\ion O{ii}] $\lambda$
3727 \AA ~\ implies that it is a good SFR {indicator} if it is
calibrated with \ion H {ii} regions older than 4.5 Ma, and used on
objects dominated by such \ion H {ii} regions. As the mass dependence
of [\ion O{ii}] $\lambda$ 3727 \AA ~\ is similar to the mass
dependence of [\ion N {ii}] $\lambda$ 6584 \AA, the intensity ratio of
these lines has a small mass dependence. This supports the use of that
ratio as an abundance {indicator}. We also check that the R$_{23}$
parameter is almost independent of mass, while S$_{23}$ and
$\eta^\prime$ have a non-trivial mass dependence.

The estimation methods that use mass independent {indicators} can be
calibrated with no consideration for the mass of the objects. If the
{indicator} is mass dependent, however, the estimation method should
be recalibrated taking into account the ionizing cluster mass. For
those methods, semi-empirical calibrations are, at least for now, more
appropriate.

An additional result obtained from our models is an intrinsic limit of
about $\Delta E(B-V)$ = 0.1 in the accuracy of the extinction
estimations carried out under the hypothesis assumed in synthesis
models (case B and $T_{\mathrm{e}}$= 10$^4$ K). This error is
diminished if tailored models are used.

Finally, we have shown that the relevant collision lines have a non
linear dependence with the cluster mass, and that, in general, not all
the collisional lines have the same dependence (slopes in the $\log M
- \log L$ differs from line to line). A direct implication of this
result is that different collisional lines in the emission line
spectrum of a non-active galaxy have different contributions from
different H{\sc ii} regions and, hence, it depends on the H{\sc ii}
population in the galaxy (and the underlying initial mass cluster
distribution) and that low mass clusters ($< 10^4$ M$_\odot$) can
contribute more than 50\% of the luminosity of the line. In this
situation, any attempt to infer physical properties from the emission
line spectrum of a non-active galaxy by comparison with single nebulae
photoionization models and without taking into account the IMF
sampling effects may be a risky business.

\begin{acknowledgements}
  We thank the referee B. Groves for his patient and his comments,
  which had improve the paper. This work has been supported by the
  Spanish {\it Programa Nacional de Astronom\'\i a y Astrof\'\i sica}
  through FEDER funding of the project AYA2004-02703 and
  AYA2007-64712.

\end{acknowledgements}

\end{document}